\documentclass[fleqn,
usenatbib]{mnras}

\usepackage{newtxtext,newtxmath}
\usepackage{anyfontsize}

\usepackage[T1]{fontenc}

\usepackage{graphicx}	
\usepackage{amsmath}	
\usepackage{hyperref}
\usepackage[normalem]{ulem}
\usepackage[dvipsnames]{xcolor}

\usepackage{tikz}

\graphicspath{{figures/}} 

\usepackage{lineno}

\defcitealias{james+21}{J21}
\newcommand{\JJ}{\citetalias{james+21}}
\newcommand{\VICE}{\textsc{vice}}

\newcommand{\fruity}{\texttt{\hyperlink{fruity}{FRUITY}}}
\newcommand{\nugrid}{\texttt{\hyperlink{nugrid}{NuGrid}}}
\newcommand{\monash}{\texttt{\hyperlink{monash}{Monash}}}
\newcommand{\aton}{\texttt{\hyperlink{aton}{ATON}}}
\newcommand{\cfactor}{2.14}
\newcommand{\nsubgiants}{14,066}

\newcommand{\caah}{[C/Mg]-[Mg/H]}
\newcommand{\caafe}{[C/Mg]-[Mg/Fe]}
\newcommand{\co}{{\rm [C/O]}}
\newcommand{\Hii}{H~{\sc ii}}

\newcommand{\Yct}{{y_{\rm C}^{\rm  tot}}}
\newcommand{\Ycc}{{y_{\rm C}^{\rm CC}}}
\newcommand{\Ymg}{{y_{\rm Mg}^{\rm CC}}}
\newcommand{\Ycagb}{{y_{\rm C}^{\rm AGB}}}
\newcommand{\ycagb}{\y_{\rm C}^{\rm AGB}}
\newcommand{\y}{p}

\newcommand{\aagb}{\beta_{\rm C}^{\rm AGB}}
\newcommand{\fagb}{f_{\rm C}^{\rm AGB}}
\newcommand{\yo}{y_0}
\newcommand{\yl}{y_{\rm low}}
\newcommand{\zcc}{\zeta_{\rm C}^{\rm CC}}

\newcommand{\Mo}{ {\rm M}_{\sun}}
\newcommand{\Zo}{ Z_{\sun}}

\newcommand{\about}[1]{${\sim} #1$}

\makeatletter
\DeclareRobustCommand\citepos
  {\begingroup
   \let\NAT@nmfmt\NAT@posfmt
   \NAT@swafalse\let\NAT@ctype\z@\NAT@partrue
   \@ifstar{\NAT@fulltrue\NAT@citetp}{\NAT@fullfalse\NAT@citetp}}

\let\NAT@orig@nmfmt\NAT@nmfmt
\def\NAT@posfmt#1{\NAT@orig@nmfmt{#1's}}

\makeatother

\title[The origin and galactic evolution of carbon]{The galactic chemical evolution of carbon: Implications for stellar nucleosynthesis}

\author[D. A. Boyea et. al.]{%
Daniel A. Boyea,$^{1, 2, 3}$\thanks{%
Contact e-mail:~\href{mailto:danielaboyea@gmail.com}{danielaboyea@gmail.com}}
James W. Johnson,$^{4, 2, 3}$
and
David H. Weinberg$^{2,3}$
\\
$^{1}$Department of Physics \& Astronomy, University of Victoria, 3800 Finnerty Road, Victoria, BC, V8P 5C2, Canada
\\
$^{2}$Department of Astronomy, The Ohio State University, 140 W. 18th Ave., Columbus, OH, 43210, USA
\\
$^{3}$Center for Cosmology \& Astroparticle Physics (CCAPP), The Ohio State University, 191 W. Woodruff Ave., Columbus, OH, 43210, USA\\
$^{4}$The Observatories of the Carnegie Institution for Science, 813 Santa Barbara St., Pasadena, CA, 91101, USA
}

\date{Accepted XXX. Received YYY; in original form ZZZ}

\pubyear{2025}

\begin{document}
\label{firstpage}
\pagerange{\pageref{firstpage}--\pageref{lastpage}}
\maketitle

\begin{abstract}
Carbon (C) is thought to be produced by both core collapse supernovae (CCSN) and asymptotic giant branch (AGB) stars, but the relative contributions of these two sources are uncertain. 
We investigate the astrophysical origin of C using models of Galactic chemical evolution (GCE) appropriate for the Milky Way disk.
We benchmark our results against APOGEE subgiant abundances.
The trend between [C/Mg] and [Mg/H] is set by the total C yield as a function of metallicity. Observations indicate a gently rising [C/Mg] with [Mg/H], but AGB C production is predicted to decline with metallicity. 
Our sample therefore favours a scenario in which CCSN yields rise with metallicity to offset declining AGB C yields and drive a subtle increase in [C/Mg] with [Mg/H].
This result is consistent with massive star nucleosynthesis models incorporating rotation.
The [C/Mg]-[Mg/Fe] trend is sensitive to delayed enrichment and therefore constrains the amount of AGB C production.
Given the slope of this relation, we find that AGB stars likely account for 10--40 per cent of C at solar metallicity.
Artificially shifting the AGB C yields towards lower mass stars with longer lifetimes also improves agreement with the observed [C/Mg]-[Mg/Fe] trend, possibly indicating a discrepancy with stellar evolution predictions or our assumed Fe production rate.
\end{abstract}




\section{Introduction}

\begin{figure*}
    \centering
    \includegraphics{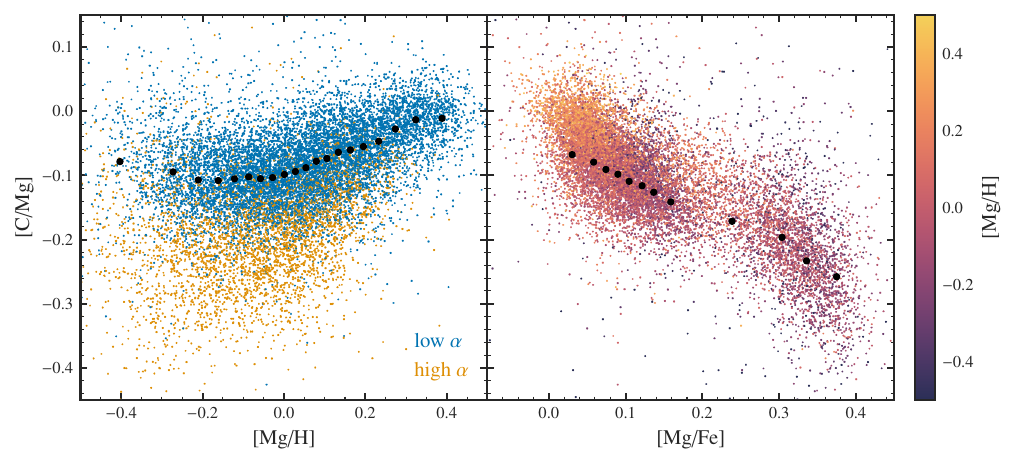}
    \caption{The [C/Mg] ratio versus [Mg/H] (left) and [Mg/Fe] (right) for the \citet{jack}~sample of APOGEE subgiants. {\bf Left:} High- and low-$\alpha$ stars are shown in blue and orange, respectively, using the separation defined in Eq.~\ref{eq:high_alpha}. The black points represent the median trend along the low-$\alpha$ sequence. {\bf Right:} Stars are colour-coded by their [Mg/H] abundance. The black points show the median [C/Mg]-[Mg/Fe] sequence for stars in the $-0.15 \leq [{\rm Mg/H}] \leq -0.05$ range. We use these median trends as our empirical benchmark in subsequent figures. } \label{fig:subgiants}
\end{figure*}

Despite being the second most abundant metal in the Universe~\citep[e.g.,][]{magg+22}, the nucleosynthetic origin of C is poorly understood. 
There is broad agreement that both massive and asymptotic giant branch (AGB) stars produce C \citep[e.g.,][]{jennifer19}, but their relative yields are disputed (see, e.g., the reviews by \citealt{romano22} and \citealt{RM21}). 
From a Galactic chemical evolution (GCE) perspective, some authors have argued that massive stars should dominate~\citep[e.g.,][]{prantzos+94, HEK00, romano+20, franchini+20, gustafsson22}, and others have argued that AGB stars should contribute at least half of C in the Universe \citep[e.g.,][]{tinsley79, chiappini+03, carigi+05, cescutti+2009, mattsson10, KKU11, KKL20}. In addition, some studies like \citet{rybizki+17} find that AGB stars may produce about 20 to 50 per cent of solar C depending on the yield tables assumed.
In this paper, we aim to constrain C production by comparing abundance measurements in subgiant stars with multi-zone GCE models.%

Predictions of C yields from stellar evolution and supernova (SN) models are uncertain for numerous reasons (see, e.g., the reviews by \citealt{romano+10, KL14}).
Some prominent unknowns include mass loss rates \citep{sukhbold+16, beasor+2020}, rotational mixing \citep{frischknecht+16, LC18}, nuclear reaction rates \citep{herwig+05, herwig+austin2004}, and convection \citep{chieffi2001, ventura+13, issa+25}.
Recent work additionally suggests that small changes in a star's initial mass of 
${\sim}0.01\,\Mo$  have significant effects on the evolution and explosion outcomes of massive stars \citep{bruenn+2023, vartanyan_burrows2023}. Most yield tables have far coarser initial-mass resolution.  Furthermore, many stars are found in binary systems \citep{sana+12}, the effects of which are almost entirely
unexplored in stellar nucleosynthesis models \citep{farmer+21}.
Robust predictions of metal yields from stellar evolution models therefore remain elusive.

Given the challenge of theoretical yield predictions, previous work has attempted to constrain chemical yields empirically. 
Even early chemical evolution work, such as \citet{garnett1997}, began to test stellar theory with observed abundance trends.
More recently, \citet{weinberg+19, weinberg+22} analysed the abundance ratio trends of high- and low-$\alpha$ sequence stars in APOGEE~\citep{apogee17} to infer relative yields of prompt and delayed enrichment sources and their metallicity dependence.
\citet{emily+19, emily+22, emily+24} applied the same methodology to GALAH~\citep{DeSilva2015, Martell2017}. In addition, models such as \citet{maoz+graur2017}, \citet{eitner+2020}, and \citet{dubay+24} have used abundance trends of Fe and related elements to understand the physics and origin of SN Ia.
Most directly relevant to our analysis, \citet{james+24} applied Milky Way chemical evolution models to N and recommended a yield prescription that reproduces the observed relation between N and O abundances. We  extend their methodology to C in this paper.


We use a sample of subgiant stars from APOGEE \citep{apogee17} as our primary observational constraint.
Subgiants' photospheric C abundance closely matches their birth composition \citep{gilroy89, korn+07, lind+08, souto+18, souto19}.
In other evolutionary phases, the measured abundances are known to be affected by internal processes (see discussion in Section~\ref{sec:data_selection} below).
Our sample should therefore be unaffected by uncertainties in inferring birth abundances.

In Section~\ref{sec:data_selection} we describe the selection for our subgiant observational sample.
In Section~\ref{sec:nucleosynthesis}, we discuss theoretical estimates of AGB and CCSN C yields.
In Section~\ref{sec:vice}, we describe the details of our GCE models.
In Section~\ref{sec:results}, we present our model predictions and discuss their observationally distinguishable predictions. 
We discuss our conclusions in Section~\ref{sec:conclusions}.

\section{The Subgiant Sample}\label{sec:data_selection}

We use subgiant stars as our empirical benchmark.
Subgiants have the advantage that their photospheric C and N abundances should closely approximate their birth mixture.
During main sequence evolution, metals can fall below the convective envelope (\textit{gravitational settling}; e.g., \citealt{turncotte+98}).
When stars evolve off the main sequence and become subgiants, these metals are reincorporated into the convective envelope \citep[]{gratton+00, souto19}. 
However, once a star becomes a red giant, \textit{first dredge up} pollutes its photosphere with CNO-processed material from the core \citep{iben67, KL14}.

We use the sample of APOGEE DR17 subgiants from \citet{jack}. We select stars from the {\sc aspcap} pipeline (revision 1) within the $\log g$-$T_\text{eff}$ polygon given by
 \begin{equation}
    \begin{cases} \label{eq:subgiant_selection}
        \log \text{g} \geq 3.5 \\
        \log \text{g} \leq 0.004\,T_{\rm eff} - 15.7 \\
        \log \text{g} \leq 0.0007\,T_{\rm eff} + 0.36 \\
        \log \text{g} \leq -0.0015\,T_{\rm eff} + 12.05 \\
        \log \text{g} \geq 0.0012\,T_{\rm eff} - 2.8. \\
    \end{cases}
\end{equation}
Following \citet{jack}, we also exclude stars with the flags:
        \verb|APOGEE_MIRCLUSTER_STAR|,
        \verb|APOGEE_EMISSION_STAR|,
        \verb|APOGEE_EMBEDDEDCLUSTER_STAR|,
        \verb|APOGEE2_YOUNG_CLUSTER|,
        \verb|APOGEE2_W345|,
        \verb|APOGEE2_EB|, and
        \verb|DUPLICATE|.
We additionally remove stars with no reported (or flagged unreliable) C, N, Mg, or Fe abundances. These cuts yield a sample containing \nsubgiants\ subgiants.

Fig.~\ref{fig:subgiants} shows the [C/Mg] abundance ratios of our subgiant sample as a function of [Mg/H] and [Mg/Fe].\footnotemark{} In this sample, [C/Mg] increases with [Mg/H] and decreases with [Mg/Fe] at fixed [Mg/H].
Some previous studies \citep[e.g.,][]{tinsley79, KKU11, KKL20} have focused primarily on the [C/Fe]-[Fe/H] ratio, interpreting an increase or decrease in [C/Fe] with [Fe/H] as an indication of the timescales of C production by a single stellar population relative to Fe. However, we show in Section~\ref{sec:n_and_fe} below that the metallicity dependence of C yields produces similar effects as delayed enrichment sources like SNe Ia and AGB stars \citep[see also][]{romano22}. Our inclusion of Mg allows us to more confidently pin down the characteristic delay-time of C production by AGB stars, and by extension the metallicity-dependence of its yield, by connecting C abundances to the decline in [Mg/Fe] with time.
Following \citet[eq.~1]{jack}, low-$\alpha$ stars are defined to have 
\begin{equation}\label{eq:high_alpha}
\begin{cases}
\text{[Mg/Fe]} <0.16-0.13\,\text{[Fe/H]}, & \text{[Fe/H]}<0\\
\text{[Mg/Fe]} <0.16, & \text{[Fe/H]}>0. \\
\end{cases}
\end{equation}
Compared to the low-$\alpha$ sequence in \caah, the high-$\alpha$ sequence has a systematically lower [C/Mg] value and steeper slope. This is consistent with the shape of the \caafe{} trend.
Note that the \caafe{} trend follows a two-component linear distribution, with a separation in [Mg/Fe] corresponding to the gap between high- and low-$\alpha$ stars. Additionally, the \caafe{} trend is approximately independent of metallicity. We aim to reproduce these trends in our GCE models.

\footnotetext{In this paper, we use the standard notation for chemical abundances. $[A/B] = \log_{10}\left(A/B\right) - \log_{10}\left(A_{\sun}/B_{\sun}\right)$, i.e. $[A/B]$ is the logarithm of the mass ratio between A and B, scaled such that $[A/B]=0$ for the Sun. (See Table~\ref{tab:fiducial_mod} for the solar scale).
}

While any abundance trend likely suffers from systematic biases, we find that these biases do not impact our qualitative conclusions. 
We have also considered subgiants and dwarfs with similar selection criteria from GALAH DR4 \citep{galah_dr4}, Gaia-ESO DR5.1 \citep{gaiaeso_dr5.1}, LAMOST \citep{lamost}, and mixing-corrected APOGEE giants \citep{vincenzo+21}. 
All of these samples show similar trends in \caah\ and \caafe{}, but with global systematic offsets of ${\sim} 0.1\,{\rm dex}$ and slight differences in the slope of each trend. Global offsets would affect our inferred total C trend with metallicity. However, given that the \caafe{} trend is similar among different surveys, we would recover similar results for the inferred relative contribution of AGB stars to C production. Our results should therefore be robust against systematic differences between these surveys.

\section{Carbon Nucleosynthesis}\label{sec:nucleosynthesis}
Here, we review the processes that produce C in both high- and low-mass stars. We define yields as used in this paper and discuss the physics, predictions, and uncertainties from stellar model predictions. Our yield assumptions for other elements are presented in Section~\ref{sec:vice}. 

A yield quantifies the new production of a chemical species. For a given star, we define the {\it stellar yield of $X$} as the fraction of a star's initial mass synthesized into an element $X$ and released.  Specifically, we use {\it net} yields here, which subtracts the birth abundance of a chemical element from the ejected material. Net yields may become negative if more $X$ is transformed into other elements or locked inside the remnant than is produced. In contrast, a {\it gross} yield is the total mass of an element ejected over the star's lifetime and is always non-negative.
 We also describe yields by their population-averaged quantities, or {\it integrated yields}, which refer to the amount of newly produced $X$ per unit mass of star formation. These yields are calculated from the stellar mass-weighted average of net stellar yields over the initial mass function (IMF).

\subsection{Low and intermediate mass stars}\label{sec:agb}

\begin{figure*}
    \centering
    \includegraphics[]{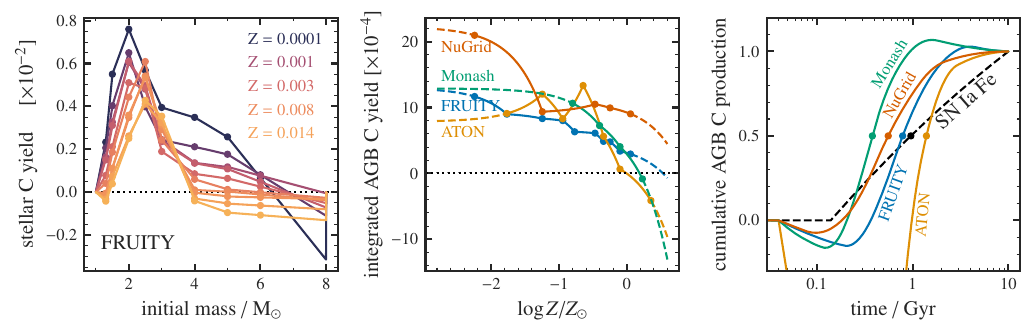}

    \caption[]{
        AGB yields as functions of mass (left), metallicity (middle), and delay-time (right). \textbf{Left:} The net stellar C yield for the \fruity{} AGB yield table as a function of initial mass and colour-coded by metallicity.
        \textbf{Middle:} The integrated C yield from AGB stars, $\Ycagb$, as a function of metallicity for each yield table,  calculated $10\,{\rm Gyr}$ after a stellar population forms. (Note that the shape of the integrated yields arises from our linear interpolation of yields in $Z$. Changing to linear interpolation in $\log Z$ does not substantially affect our results.)
        \textbf{Right:} Cumulative C production as a function of age for a single stellar population of metallicity $\log Z/\Zo=-0.1$. The dashed black line shows the cumulative return fraction of Type Ia supernovae ($\propto t^{-1.1}$) for comparison. The cumulative production of \aton{} reaches a minimum of -3 at a time of 0.3 Gyr.
}

   \label{fig:agb_yields}

\end{figure*}

An asymptotic giant branch (AGB) star is a low to intermediate mass star undergoing He shell burning, its last nuclear burning phase (see e.g., \citealt{PR2023}). AGB stars enrich the interstellar medium (ISM) by expelling processed material during thermal pulses.

We explore four different AGB star yield tables from the literature, which provide well-sampled grids in mass and metallicity. We refer to the tables as 
\begin{description}
    \item \hypertarget{fruity}{\texttt{FRUITY}}: \citet{cristallo+11, cristallo+15}
    \item \hypertarget{aton}{\texttt{ATON}}: \citet{ventura+13,ventura+14,ventura+18, ventura+20}
    \item \hypertarget{monash}{\texttt{Monash}}: \citet{KL16, karakas+18}
    \item \hypertarget{nugrid}{\texttt{NuGrid}}: \citet{pignatari+16, ritter+18, battino+19, battino+21}
\end{description}
Each AGB yield table reports yields for stellar models on a grid of masses and metallicities as reported in Table~\ref{tab:agb}.
We use the same tables as \citet{james+23}, except that we have swapped \citet{karakas10} for \citet{pignatari+16}. (See Appendix~\ref{sec:extra_agb} for the implementation of these yields and discussion of the differences between AGB yield models.)

The left panel of Fig.~\ref{fig:agb_yields} shows the C yields for the \fruity{} table (see Appendix~\ref{sec:extra_agb} for plots of the other yields). C production peaks between masses of $\sim$ 2--4 $\Mo$. As metallicity increases, the net yield decreases and the mass of peak C production increases slightly. As a result, most AGB models predict that C production becomes less efficient but more rapid with increasing metallicity.

To integrate AGB yields with \VICE, we linearly interpolate yield tables in both mass and metallicity. Below the smallest mass in a table, $\ycagb$ is linearly extrapolated to 0 at $1\,\Mo$ to avoid numerical artifacts. The population-averaged AGB star yield is calculated with 
\begin{equation} \label{eq:imf-agb-yield}
    y_{X}^{\rm AGB}(Z,\tau) = 
    \int^{u_{\rm AGB}}_{m_\textrm{post-AGB}(\tau)} 
    \y_{\rm X}(m, Z)\ 
    \frac{1}{A}\frac{dN}{dm}\ m\ dm ,
\end{equation}
where $u_{\rm AGB}=8\,\Mo$ is the maximum birth mass of an AGB star, $m_\textrm{post-AGB}(\tau)$ is the birth mass of stars which complete AGB evolution at age $\tau$, and $dN/dm$ is the initial mass function (IMF). The IMF normalization is $
    A := \int_l^u m \frac{dN}{dm} dm
$
where  $l=0.08\,\Mo$ and $u = 100\,\Mo$ are the assumed minimum and maximum mass of star formation. We assume a \citet{larson74} main sequence mass-lifetime-relationship. 
Although there are other, more recent forms of the mass-lifetime relation available in the literature \citep[e.g.,][]{hurley+2000, vincenzo+2016}, the updates have primarily affected the low- and high-mass ends of the IMF, leaving the AGB star mass range (${\sim}1$--$8\,\Mo$) largely unaffected. The highest mass stars live sufficiently short lives that their enrichment can be approximated as instantaneous for most purposes \citep[see discussion in, e.g.,][]{WAF17}, while the lowest mass stars have lifetimes larger than the age of the Universe and therefore do not meaningfully influence chemical evolution. As a consequence, swapping our mass-lifetime relation for any of these alternatives has no impact on our conclusions. 
We take the total stellar lifetime (through post-AGB) to be $1.1\times$ the main sequence lifetime. Thus, the mass $m_\textrm{post-AGB}$ is slightly larger than the main-sequence turn-off mass.

The IMF-averaged yields computed from the literature tables differ in normalization and metallicity dependence.  
The middle panel of Fig.~\ref{fig:agb_yields} shows IMF-averaged C yields for each AGB model as a function of metallicity, computed at time $\tau = 10$ Gyr after the formation of a single stellar population.
The normalizations span a factor of \about{2} (e.g., between $6\times 10^{-4}$ and $12 \times 10^{-4}$ at $Z=0.33\,\Zo$).
Additionally, the slope varies by a factor of ${\sim}3$.
Both \nugrid{} and \aton{} predict slightly non-monotonic C yields with $Z$, but modelling systematics may exceed the scale of non-monotonic features.
These variations are due to different choices of reaction rates, convection treatments, and mass-loss rates (see discussion below and \citealt{james+23}).

The right panel of Fig.~\ref{fig:agb_yields} shows the time-dependence of C production by AGB stars in a single stellar population of metallicity ${\rm [M/H]} = -0.1$ as a function of age. 
As the mass range $2\,\Mo\lesssim M \lesssim 4\,\Mo$ is predicted to be most important for C production, half the yield is produced before \about{1}\,Gyr, similar to Fe production by SN Ia. 
\monash{} weights C production more heavily towards high-mass AGB stars, resulting in shorter delay times, whereas the \fruity\ model predicts a slightly longer time-scale of \about{1}\,Gyr. In any case, little to no C is produced more than 2\,Gyr after a star formation event. Fe production, in contrast, continues steadily for 10\,Gyr. 

In AGB stars, {third dredge up} (TDU) and {hot bottom burning} (HBB) are two competing processes driving C evolution.
TDU accompanies thermal pulses, where material from the CO core is mixed with the envelope, increasing surface C abundances 
later released to the ISM \citep{KL14}.  
HBB\ is the activation of the CNO cycle at the bottom of the convective envelope, which converts most C into N.
TDU increases C yields, while HBB decreases C. 


Both HBB and TDU result in mass and metallicity-dependent C yields. 
Lower mass stars ($\lesssim 1\,\Mo$)  do not experience (strong) TDU. As a result, C yields from these stars are affected only by first dredge-up, resulting in small net C yields \citep{karakas10}.
Above ${\sim}1\,\Mo$, TDU becomes more important, enriching the outer layers with C.
In AGB stars more massive than ${\sim}5\,\Mo$, efficient HBB turns most of the $^{12}$C into $^{14}$N.
TDU is more efficient for compact, massive cores. HBB requires a more massive AGB star to reach sufficient temperatures at the base of the convective envelope to initiate the CNO cycle. Both processes are more efficient at low metallicity due to the lower opacities, hotter temperatures, and more compact internal structures. The details of each process are ultimately set by uncertain choices in stellar evolution models. 


The primary uncertainties in AGB stellar evolution are convection, nuclear reaction rates, and mass loss prescriptions.
\citet{ventura+15} compared the \aton{} and \monash{} models, finding that differences in yields are primarily driven by differences in mass-loss and convection prescriptions. \aton{} predicts strong HBB to set in ${\sim}1\,\Mo$ lower than in \monash{}, leading to much lower C yields.
Moreover, the $^{14}{\rm N}({\rm p},\gamma)^{15}{\rm O}$ reaction rate uncertainty causes a factor of 2 difference in predicted C yields \citep{herwig+austin2004, HAL2006}.
Mass-loss rates, convection, and extra mixing are not sufficiently well understood to be computed from first principles. Thus, most of the physics included in stellar modelling is empirically calibrated. 

Super AGB stars are the class of AGB stars with masses ${\sim}8$ to ${\sim}10\,\Mo$, which burn O into Si and Mg. While we neglect super-AGB stars in the discussion above, these stars are not predicted to be substantial producers of C, similar to high-mass AGB stars. From the \citet{doherty+14, doherty+14b} yield table (extending the \monash\ yields), the highest net fractional yield is only $-0.002$ for super solar metallicity, which would have a negligible impact on the integrated yield. Additionally, as discussed in Section~\ref{sec:agb_results} below, the short lifetimes of high-mass AGB and super-AGB stars imply that they minimally affect our conclusions and can be grouped with CCSN yields.

For our models to better match observations, we uniformly amplify the yield tables according to
\begin{equation} \label{eq:alpha}
        \ycagb(m, Z) \rightarrow \aagb\ \ycagb(m, Z).
\end{equation}
We also define the AGB yield fraction, 
\begin{equation}\label{eq:f_agb}
    \fagb := \frac{\Ycagb(\Zo)}{\Ycc(\Zo) + \Ycagb(\Zo)}.
    \end{equation}
This quantity should approximately represent the fraction of the Sun's birth C abundance originating from AGB stars. However, since the Sun was not enriched entirely by solar metallicity predecessors, the equivalence of $\fagb{}$ to the yield ratio is not exact.
As our fiducial AGB yield, we use the \fruity\ table with $\aagb=\cfactor$ or equivalently $\fagb=0.25$ based on a best-fit computed in Appendix~\ref{sec:extra_mcmc}.

\begin{table*}
	\centering
    \caption[]{For each AGB yield set, the IMF-averaged AGB C yield at solar metallicity $y_{\rm C, 0}^{\rm AGB}$, the fraction of solar C produced in the model $f_\odot^{\rm AGB}$, and the masses and metallicities each yield table is sampled on.
    $f_{\sun}^{\rm AGB}$ is calculated based on $y_{\rm C, \sun}^{\rm AGB}$ assuming the fiducial total C yield of $\Yct = 26.7\times 10^{-4}$ and $\aagb=1$.
    }

	\label{tab:agb}
    \begin{tabular}{c  c  c p{4cm} p{4cm}} 
		\hline 
        AGB table 
                & $y_{\rm C, \sun}^{\rm AGB}\times10^4$ 
                &  $f_{\sun}^{\rm AGB}$
                & masses ($\Mo$) & metallicities ($Z$)\\
        \hline
        \fruity 
                & 3.11
                & 0.12
                & 1.3, 1.5, 2, 2.5, 3, 4, 5, 6
                & 0.0001, 0.0003, 0.001, 0.002, 0.003, 0.006, 0.008, 0.01, 0.014, 0.02
                \\
        \aton 
                & -0.009
                & -0.0003
                & 1.5, 2, 2.5, 3, 3.5, 4, 4.5, 5, 6, 6.5, 7
                & 0.0003, 0.001, 0.002, 0.004, 0.008, 0.014, 0.04
                \\
        \monash 
                &  2.96
                & 0.11& 1, 1.25, 1.5, 1.75, 2.25, 2.5, 2.75, 3, 3.25, 3.5, 3.75, 4, 4.5, 5, 5.5, 6, 7 
                & 0.0028, 0.007, 0.014, 0.03
                \\
        \nugrid 
                & 9.25
                &  0.35
                & 1, 1.65, 2, 3, 4, 5, 6, 7
                &  0.0001, 0.001, 0.006, 0.01, 0.02
                \\
		\hline
	\end{tabular}
\end{table*}

\subsection{Massive stars}

Massive stars form $^{12}$C in their cores through the triple-$\alpha$ reaction (He-burning), which is later released to the ISM through winds and supernovae. Both material ejected through stellar winds and through supernovae contribute to C production, but we generally use the term CCSN to refer to the sum of both modes of massive star enrichment.

We approximate massive star enrichment as occurring instantaneously after a star formation event. Under this approximation, the yield is a constant  of proportionality between the metal production rate and the star formation rate (SFR):
\begin{equation}
    \dot{M}_X^{\rm CC} = y_{X}^{\rm CC}\, \dot{M}_\star.
\end{equation}
The CCSN integrated yield is given by
\begin{equation}
y_{X}^{\rm CC} = \int_{l_{\rm CC}}^{u} \left[E(m) m_{X\rm, SN} + w_X - Z_X\,(m-m_{\rm rem})\right]\,\frac{1}{A} \frac{dN}{dm}\ dm,
\end{equation}
where $E(m)$ is the fraction of stars of mass $m$ which explode, $m_{X,\rm SN}$ is the mass of $X$ ejected during the CCSN, $w_X$ is the mass of $X$ ejected through stellar winds, $Z_X$ is the initial abundance of $X$, $m_{\rm rem}$ is the remnant mass, $dN/dm$ is the IMF, $l_{\rm CC}=8\,\Mo$ is the minimum initial mass of a CCSN progenitor, and $u$ is the maximum mass in the CCSN yield table. 

\begin{figure}
    \centering
    \includegraphics{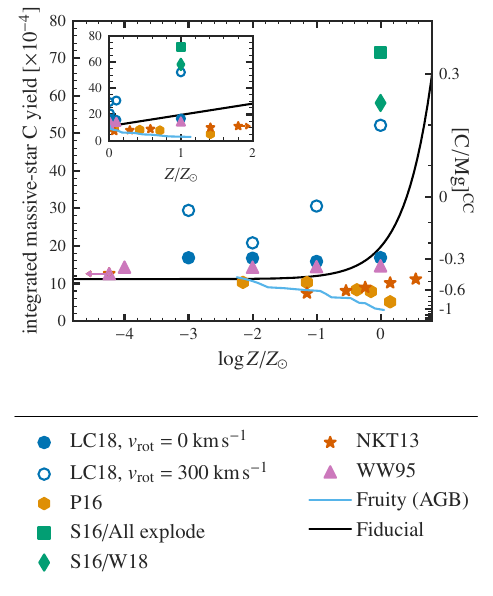}
    \caption[]{
        IMF-integrated C yields from massive stars plotted as a function of metallicity.
        The right axis provides the equivalent CCSN [C/Mg]  ratio, assuming our fiducial $y_{\rm Mg}$ yield. The black line is our fiducial massive star yield (see Eq.~\ref{eq:y_cc}).
        Yields are shown for tables from \citet[LC18, blue circles]{LC18}, \nugrid{} (P16, orange hexagons),  \citet[S16, green square and diamond]{sukhbold+16},  \citet[NKT13, red stars]{NKT13}, and 
    \citet[WW95, pink triangles]{WW95}. \citet{sukhbold+16} report yields for different black hole landscapes, while \citet{LC18} provide yields at different rotational velocities.
   The light blue line denotes $\Ycagb$ from the \fruity{} AGB star model for comparison (see Fig.~\ref{fig:agb_yields}). All models include wind yields. 
   The inset shows the same comparison except with a linear-scaled metallicity axis
}
    \label{fig:y_cc}
\end{figure}

CCSN models predict a wide range of C yields, spanning nearly a factor of ten, even at fixed metallicity.
Fig.~\ref{fig:y_cc} shows the integrated yields of massive star models from the literature. At solar metallicity, \citepos{sukhbold+16} C yield is $\approx0.5$ dex higher than \citepos{LC18} non-rotating models, likely driven by much stronger (C-enhanced) mass loss in \citet{sukhbold+16} for stars more massive than $\gtrsim40\,\Mo$.
 The \cite{LC18} models including rotation show that the induced mixing (e.g., \citealt{frischknecht+16}) can dramatically increase the magnitude and metallicity dependence of $\Ycc$. As we will later empirically show, CCSN C production needs to be strongly metallicity-dependent at $Z \approx \Zo$, which is consistent with \citepos{LC18} rapidly rotating models and to a lesser extent \citet{NKT13}.
Fig.~\ref{fig:agb_yields} also shows the IMF-averaged AGB star yields from the \fruity\ model. Especially at $Z\approx \Zo$, most CCSN models dominate AGB C production.

If we consider CCSN production alone, the [C/Mg] ratio depends only on the yield ratio,
\begin{equation}\label{eq:c_mg_cc}
    {\rm [C/Mg]^{CC}} = \log_{10}\left( \frac{\Ycc}{\Ymg}\right) - \log_{10} \left( \frac{Z_{{\rm C},\ \sun }}{Z_{{\rm Mg},\ \sun }} \right).
\end{equation}
Because the yield ratio depends on metallicity, the abundance ratio will settle to an equilibrium that depends only on the equilibrium metallicity \citep{WAF17, james+23}. The right axis of Fig.~\ref{fig:y_cc} shows the  ${\rm [C/Mg]^{CC}}$ ratio for different CCSN C yields and our fiducial Mg yield. Different models have values of ${\rm [C/Mg]^{CC}}$ between $ -0.8$ and $+0.3$ at near-solar metallicity, illustrating the wide range of CCSN predictions.

The metallicity dependence of CCSN C yields is due to an interplay between rotation, opacity, and stellar evolution. As opacity increases with metallicity, massive stars experience stronger winds. In particular, C-enriched envelope material is lost through winds before conversion into heavier elements, resulting in metallicity-dependent C production \citep{LC18}.
While there are many stellar models providing predictions of CCSN yields, the results of these models are highly uncertain. Furthermore, rotation, binarity, and explodability introduce substantial uncertainties in CCSN predictions \citep{farmer+21}.

Given the wide range of predictions for CCSN C yields, we choose to adopt a linear yield model instead of a particular literature yield table. We parameterize the yield as
\begin{equation}\label{eq:y_cc}
    \Ycc =  \yo + \zcc \left(\frac{Z - \Zo}{\Zo}\right),
\end{equation}
where $\yo$ represents the yield at solar metallicity and $\zcc$ is the slope of the metallicity dependence. 
$\yo$ can be thought of as the yield of primary CCSN C (produced without dependence on metallicity) and $\zcc$ as the normalization of secondary CCSN C (produced with a linear dependence on metallicity). 
For our fiducial model, we adopt $\yo = {19.7}\times 10^{-4}$ and $\zcc = {8.56} \times 10^{-4}$, leading to reasonable agreement with the \caah{}, \caafe{}, and gas-phase abundance relations (see Section~\ref{sec:results} and Appendix~\ref{sec:extra_gas}). The rise in $\Ycc$ with metallicity is intended to qualitatively capture the effects of rotation (see the $v_\text{rot}=300\,\text{km}\,\text{s}^{-1}$ model from \citealt{LC18} in Fig.~\ref{fig:y_cc}). However, we refrain from using a detailed match to any one rotating star model, since stars should span a range of rotational velocities.
Our linear prescription is able to capture all of the effects necessary for the purposes of this paper, despite its simplicity (see the inset of Fig.~\ref{fig:y_cc}). We have repeated our analysis with higher order prescriptions (e.g., second-, third-, and fourth-order polynomials) and found that these additional degrees of freedom do not change our qualitative conclusions. The metallicity dependence may well be more complex than Eq.~\ref{eq:y_cc}, but current observations do not provide good constraints on more complex models.

\subsection{Other sources}

SN Ia produce negligible C. Of the models from \citet{iwamoto+99}, \citet{seitenzahl+13}, and \citet{gronow+21a, gronow+21b}, the maximum C production is less than 4 per cent of our required solar C yield, and most likely less than 2 per cent (assuming  $2.2\times 10^{-3}$ SN Ia explosions per solar mass of star formation; \citealt{maoz+mannucci12}).
Novae are important contributors of $^{13}$C, but do not process enough material to impact $^{12}$C production \citep{starrfield+20}. $^{13}$C is beyond the scope of this work and accounts for a small fraction of the C in the Universe.

\section{The Multi-zone Model}\label{sec:vice}

Our models extend the \citet[hereafter \JJ]{james+21} Milky Way model. We integrate the model using the publicly available Versatile Integrator for Chemical Evolution (\VICE; \citealt{JW20}).%
    \footnote{\VICE~is available at \url{https://github.com/giganano/VICE}.}
This model is described extensively in \JJ~and summarized in \citet{james+23}. Here, we provide an overview of the relevant components.

Classical \textit{one-zone} models of chemical evolution assume instantaneous mixing of metals in the star-forming ISM\ \citep[e.g.,][]{matteucci21}. The one-zone framework is a poor approximation of the Milky Way, where the radial metallicity gradient in the ISM is believed to reflect variations in GCE parameters \citep[e.g.,][]{tinsley80, boissier+prantzos99, palla+20, james+24}. Stars can also migrate several kpc over their lifetimes, mixing populations from different chemical environments across the Galaxy \citep{sellwood+binney02, schonrich-binney09, bird+13, minchev+13}. To account for these effects, multi-zone models stitch together multiple one-zone models while mixing stellar populations between zones \citep[e.g.,][]{matteucci+francois1989}.

In these models, the Galaxy is divided into 200 rings, each representing a single 100\,pc zone. Each ring has a separate star formation rate, gas reservoir, gas inflow, and possible gas outflow. We initially assume the inside-out SFH from \JJ, where the star formation surface density $\dot{\Sigma}_\star$ is given by 
\begin{equation}\label{eq:inside_out}
    \dot{\Sigma}_\star \propto \left(1-e^{-t/\tau_{\rm rise}}\right) e^{-t/\tau_{\rm sfh}}.
\end{equation}
The rise time $\tau_\text{rise}=2$\,Gyr loosely describes when the star formation rate reaches a maximum, and $\tau_{\rm sfh}$ describes the decay time-scale of star formation, which increases with radius $R$. \JJ\ derive $\tau_{\rm sfh}(R)$ to fit the stellar age gradients measured by \citet{sanches20} from integral field spectroscopic surveys. At each $R$, the SFH is normalized to match the stellar surface density profile \citep[from][]{BHG16} assuming a total stellar mass of $5.17\times10^{10}\,\Mo$ \citep{LM15}. Star formation ends beyond a radius $R=15.5\,$kpc, but stellar populations are allowed to migrate as far as $R=20\,$kpc.  
The gas inflow is calculated to maintain the SFH for each radius and time, using an extension of the Kennicutt-Schmidt law \citep{kennicutt98} motivated by the combined observations of \citet{bigiel+10} and \citet{leroy+13} (see \JJ).
The scaling of this relationship varies with time due to the redshift dependence of $\tau_\star$ in molecular gas observed by \citet{tacconi18}.

We use the radial migration prescription described in appendix C of \citet{dubay+24} and summarized in section 3 of \citet{james+24}. The total distance each star moves is sampled from a Gaussian distribution with standard deviation $\sigma_R =  2.68\,{\rm kpc} \ (R/8\,{\rm kpc})^{0.61}\,(\tau / 8\,{\rm Gyr})^{0.33}$. Stars migrate from their initial to their final positions with a cube-root time dependence. 
These parameters arise from \citepos{dubay+24} analytic approximation of the {\tt h227} simulation \citep{bird+21, christensen+12, zolotov12, loebman12, BZ14}. 

Following \JJ{} (and more recently \citealt{james+24}), we control the slope of the radial metallicity gradient by adjusting the strength of the mass-loading in Galactic winds. The mass-loading, $\eta :=  \dot\Sigma_{\rm out} / \dot \Sigma_\star$, the ratio between the surface density rates of gas leaving the system and star formation, is set to
\begin{equation}\label{eq:mass_loading}
\eta(R) = \frac{y_{\rm Mg}^{\rm CC}}{Z_{\rm Mg}(R)} -1 + r 
\end{equation}
where 
\begin{equation}
    \log Z_{\rm Mg}(R) = \log Z_{\rm Mg,\ \sun} + 
    0.29 + 
    \begin{cases}
        -0.015(R-5) & R < 5 \\
        -0.09(R-5) & R \geq 5
    \end{cases}
\end{equation}
is adapted from \citet{hayden+14} assuming Mg is representative of $\alpha$ elements.
The recycling correction $r\approx 0.4$ accounts for the return of stellar envelopes back to the ISM (for a \citealt{kroupa01} IMF after 10 Gyr; see section 2 from \citealt{WAF17}).
This choice of $\eta(R)$ results in a [$\alpha$/H] gradient consistent with APOGEE measurements \citep[e.g.,][]{frinchaboy+13, hayden+14, weinberg+19}.
Compared to \citet{james+21, james+24}, our metallicity gradient is flatter within 5 kpc, since the outflow rates go to zero near this radius.
If we change our assumed yield scale, the values of $\eta$ will re-scale accordingly to maintain observationally-consistent chemical trends. Following \JJ, we neglect radial gas flows (see discussion below). 

To create representative stellar samples to compare with APOGEE, 
we draw \nsubgiants\ stars from the simulated stellar populations. First, we draw a random zone (weighted by the number of observed subgiants in each zone) and then draw a random particle in the zone weighted by the stellar population mass.
Galactocentric radii are taken from astroNN \citep{leung+bovy19, leung+bovy19b} to derive the number of subgiants in each zone.  The choice of stellar selection function predominantly affects the metallicity distribution, but leaves abundance trends the same in this model.

Table~\ref{tab:fiducial_mod} summarizes our yield choices. 
Our solar abundance scale is from \citet{magg+22} with a 0.04 dex gravitational settling correction. Also from \citet{magg+22}, we adopt $\Zo = 0.0176$. We use the O, Mg, and Fe yields recommended by \citet{david_fe} based on measurements of the mean CCSN Fe yield \citep{rodriguez+21, rodriguez+23} and APOGEE chemical abundances.
Our AGB N yield is scaled from \citep[eq. 14]{james+23}, based on the differences in the assumed solar N/O ratio and the median [N/Mg] or [N/O] value of each sample.  
Based on the finding by \citet{dubay+24} that GCE models favour extended delay-time distributions (DTD), we assume an event rate that is initially constant with a minimum delay time of 40 Myr, which then transitions to a $t^{-1.1}$ time-dependence after 1 Gyr. This DTD is similar to \citepos{greggio2005}  `WIDE DD' model.
We assume $1.1\times10^{-3}$ SN Ia events per unit mass of star formation  with a mean yield of $0.710\,\Mo$ of newly produced Fe per supernova \citep{david_fe}.
We use a \citet{kroupa01} initial mass function (IMF).
A known problem for CCSN yield models is that O tends to be overproduced and Mg underproduced, i.e. the O-Mg problem \citep[e.g.,][]{emily+21}. For simplicity, we adopt metallicity-independent values of O and Mg based on the observed APOGEE abundance trends from \citet{david_fe} \citep[see also][]{andrews+17, weinberg+19}. 

\begin{figure*}
\centering
\includegraphics{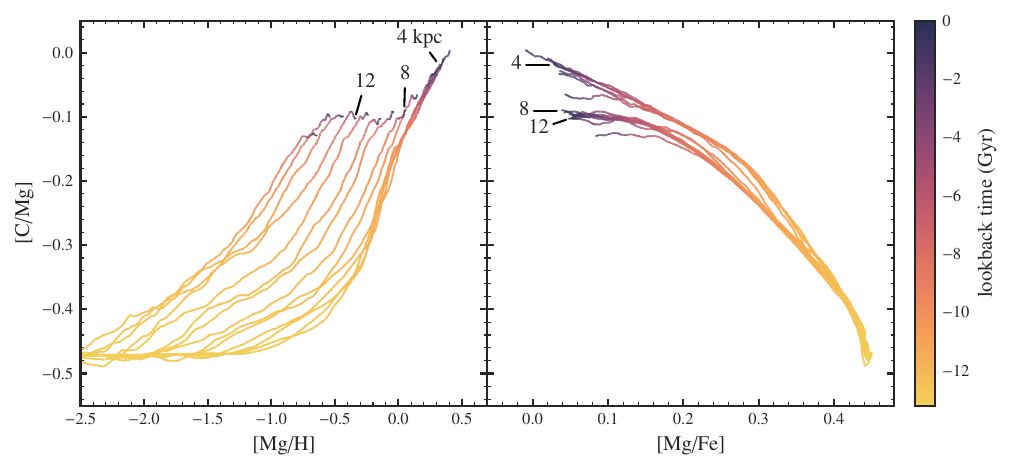}
\caption[]{
    Time evolution of gas-phase C abundances in our fiducial model for [C/Mg] versus [Mg/H] (left) and [Mg/Fe] (right).
    Each line represents a zone at a different Galactic radius, colour-coded by lookback time. We plot zones at 1 kpc intervals between 2 and 15 kpc. Evolution in [C/Mg] with [Mg/H] varies substantially between regions of the Galaxy, while evolution with [Mg/Fe] is more uniform.
}
\label{fig:c_evo}
\end{figure*}

Many GCE models with a variety of motivations omit mass loading in Galactic winds (i.e., $\eta=0$, e.g., \citealt{spitoni19, spitoni20, spitoni21, palla+20}).
This choice is based on hydrodynamic simulations in which ejected gas re-accretes to a similar location on short timescales \citep[e.g.,][]{melioli+2008, melioli+2009, spitoni+2008, spitoni+2009, oppenheimer+dave2008, hopkins+2023}. However, we emphasize that this outcome ultimately arises as a consequence of the feedback prescription built into the simulation, which is not only highly uncertain but also dependent on the implementation of the simulation itself (e.g., \citealt{li+2020, hu+2023}; see also the reviews by \citealt{veilleux+2020, thompson+heckman2024}). 
Significant mass-loading indeed arises in simulations with more energetic feedback \citep[e.g.,][]{brook+2011, gutcke+2017, nelson+2019, peschken+2021, kopenhafer+2023} or some models with cosmic ray feedback \citep[e.g.,][]{bieri+2026, hopkins+2021}. 
An alternative to outflows is to incorporate radial gas flows through the disk  \citep[e.g.,][]{lacey+fall1985, portinary+chiosi2000, spitoni+matteucci2011, bilitewski+schonrich2012, pezzulli+fraternali2016} which can also be adjusted to reproduce the ISM metallicity gradient \citep[e.g.,][]{grisoni+2018, palla+20, james2025}. We expect that models producing the same metallicity gradient with the same yields would predict similar [C/Mg] and [C/Fe] ratios regardless of whether they use outflows or radial gas flows, though we have not demonstrated this point.

\begin{table}
	\centering
    \caption[]{Solar abundance scale and fiducial yields (in units of stellar population birth mass). See Section \ref{sec:agb} for the definition of \fruity. The solar abundance scale is \citet{magg+22} corrected upwards by 0.04 dex for gravitational settling and diffusion \citep{david_fe}. For $\Ycc$, the fiducial yield is given by Eq.~\ref{eq:y_cc} with $\yo=18.7\times10^{-4}$ and $\zcc=7.9\times10^{-4}$.
    }
	\label{tab:fiducial_mod}

	\begin{tabular}{l l l l l}
		\hline
         $X$ & $Z_{X,\,\sun}$ & $y_X^{\rm cc}$ & $\y_X^{\rm agb}$ & $y_X^{\rm ia}$  \\
		\hline
        C & $33.9\times10^{-4}$ & Eq.~\ref{eq:y_cc} & $\cfactor\times$\fruity &  0 \\
        O & $73.3\times10^{-4}$ & $71.3\times10^{-4}$ & 0 & 0 \\
        Mg & $6.71\times10^{-4}$ & $6.52\times 10^{-4}$ & 0 & 0 \\
        Fe & $13.7\times10^{-4}$ & $4.73\times10^{-4}$ & 0 & $7.82\times10^{-4}$ \\
        N & $10.4\times10^{-4}$ & $5\times10^{-4}$ & $5\times10^{-4}M\left(\frac{Z}{\Zo}\right)$ & 0\\
		\hline
	\end{tabular}
\end{table}

\section{Results}\label{sec:results}

\subsection{Evolution and interpretation of carbon abundances}

Fig.~\ref{fig:c_evo} shows evolutionary tracks in \caah{} and \caafe{} in the fiducial model at 1\,kpc intervals of radius (with C yields defined in Section~\ref{sec:nucleosynthesis}).
The \caah\ trend reaches an equilibrium state within \about{5}\,Gyr. \caafe\ tracks continue to evolve due to the long tail of the SN Ia DTD.

The evolution of chemical abundances can be understood as the interplay between different processes with different time-scales. 
 Initially, CCSN dominate enrichment. The [C/Mg] abundance ratio is set by the relative yield $\Ycc/\Ymg$. Due to the metallicity dependence of $\Ycc$, [C/Mg] increases with  [Mg/H]. [Mg/H], [Fe/H], and [C/H] all rise rapidly. [Mg/Fe] is initially high due to the delayed production time-scales of SN Ia Fe, as expected from the analytic models of \citet{WAF17}.
 Of order 100 Myr later, delayed sources enrich the ISM. AGB stars release C, raising [C/Mg], and SN Ia release Fe, lowering [Mg/Fe]. 
 Several Gyr later, [Mg/H] reaches equilibrium. As the metal abundance stabilizes, so do metallicity-dependent CCSN yields, allowing C to reach equilibrium. [Fe/H] continues to rise slowly due to ongoing SN Ia from old stellar populations. 
 Finally, the system reaches approximate equilibrium and evolution slows. \
 
This evolution is predominantly driven by our chosen elemental yields, their
metallicity dependence, and their DTDs.
As we will demonstrate in Section~\ref{sec:agb_results} below, the final locus of the \caah\ trend (the end points of the curves in the left panel) depends mostly on the metallicity dependence of the total C yield. The \caafe{} trend instead reveals the nature of delayed C production. Stars with lower [Mg/Fe] values have been enriched more by SN Ia and, consequently, should also contain more delayed C from AGB stars.

\subsection{The high-mass yield slope and low-mass production fraction of C yields}\label{sec:yield_variations}
\label{sec:agb_results}

\begin{figure*}
    \includegraphics{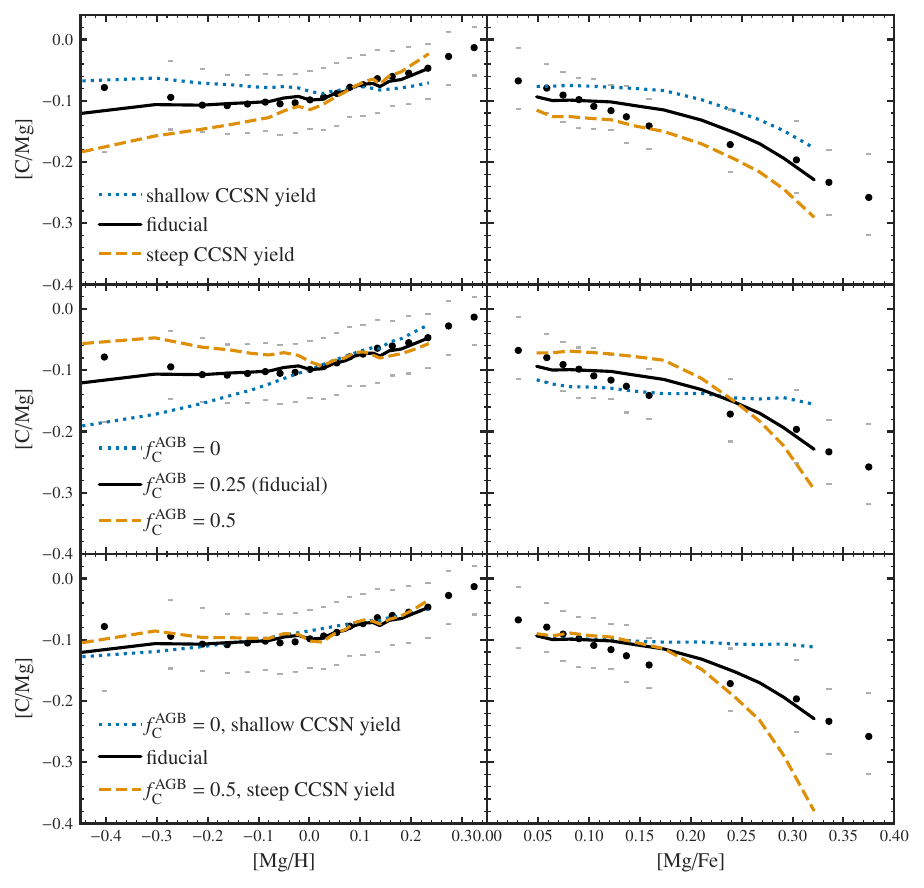}
    
    \caption[]{
        Trends in [C/Mg] with [Mg/H] (left) or [Mg/Fe] (right). 
        Stars are binned into 20 (left) or 12 (right) equal-number bins. 
        The left panel only shows low-$\alpha$ stars, and the right panel shows stars where $-0.15\leq {\rm [Mg/H]} \leq -0.05$.
        Coloured lines represent the median [C/Mg] in bins of [Mg/H] or [Mg/Fe] for each model.
        Black points and grey dashes represent the median and 16th-84th percentiles of [C/Mg] in each bin in the \citet{jack}~sample. 
        The total solar C yield is held fixed in each model. 
        \textbf{Top}: Models with different slopes of the metallicity dependence of the CCSN yield, $\zcc$ (Eq.~\ref{eq:y_cc}), where higher $\zcc$ corresponds to a more rapid increase in C production with metallicity. \textbf{Middle}: Models with different fractions of solar C produced in AGB stars, $\fagb$. 
        \textbf{Bottom:} Models where both $\fagb$ and $\zcc$ are adjusted simultaneously. 
    }
    \label{fig:zeta_f}
\end{figure*}

The top row of Fig.~\ref{fig:zeta_f} shows models with varying strengths of the $\Ycc$ metallicity dependence, $\zcc$ (see Eq.~\ref{eq:y_cc} and parameters in Table~\ref{tab:model_parameters}). With higher $\zcc$, the model predicts a steeper \caah\ trend, owing to the direct relationship between equilibrium [C/Mg] and yield ratios in Eq.~\ref{eq:c_mg_cc}. 
However, the \caafe{} trend is minimally affected by changes to $\zcc$ for stars at a given [Mg/H].
CCSN occur on much shorter time-scales than AGB and SN Ia enrichment.  The primary effect of increasing $\zcc$ in the \caafe{} plane, restricted to a metallicity slice, is to shift [C/Mg] uniformly.

The middle row of Fig.~\ref{fig:zeta_f} shows three models with different C AGB  fractions (Eq.~\ref{eq:f_agb}; parameters in Table~\ref{tab:model_parameters}).
At fixed metallicity, the \caafe\ trend is principally sensitive to how much C and Fe come from delayed sources. In the limit $\fagb \to 0$, the [C/Mg] abundance will become nearly independent of [Mg/Fe]. As $\fagb$ increases, the steepness of the \caafe{} trend increases. 

Because both $\fagb{}$ and $\zcc$ affect the \caah{} trend, it is possible to vary both of these parameters and leave the \caah{} trend nearly untouched. 
The bottom row of Fig.~\ref{fig:zeta_f} illustrates this degeneracy.
While each model has substantially different AGB contributions, ranging from $\fagb = 0$ to $0.5$, adjustments to the massive-star C yield leave the \caah{} trend mostly unaffected. While some residual differences remain, these could be addressed with a CCSN yield model that more closely compensates for the AGB yield metallicity dependence. On the other hand, the differences between these models in the \caafe{} plane are far more stark, with the trend's total amplitude ranging from almost 0 (for $\fagb{} = 0$) to nearly 0.3 dex (for $\fagb{} = 0.5$). While these trends may appear different from those in the middle panel, they are mostly only shifted vertically.

Given these models, we can interpret the physical significance of each trend. The \caah{} trend (for low-$\alpha$ stars) tells us about the total C yield. This trend is mostly independent of where C is produced at a given metallicity, but is very sensitive to changes in the total C production. Models that match \caah{} trends alone would struggle to differentiate between different production mechanisms of C. The \caafe{} trend (for a given metallicity) tells us about delayed C production. For example, the models in Fig.~\ref{fig:zeta_f} with no delayed C predict a nearly flat trend in [C/Mg] with [Mg/Fe], even with metallicity-dependent C production. As such, the \caafe{} trend is a useful diagnostic of $\fagb$.

\subsection{Iron and Nitrogen} \label{sec:n_and_fe}

\begin{figure}
    \centering
    \includegraphics{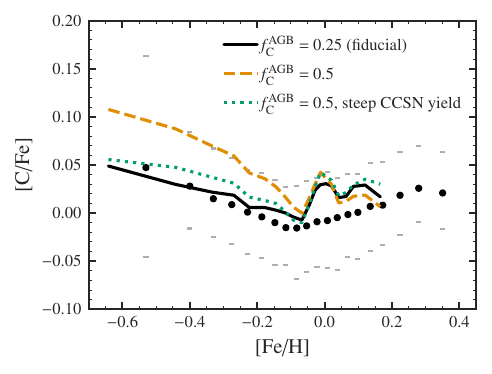}
    \caption{Similar to the left panels of Fig.~\ref{fig:zeta_f} except using Fe as the reference element instead of Mg. Increasing the fraction of AGB production of C similarly affects the [C/Fe]-[Fe/H] trend as decreasing the metallicity dependence of CCSN C production. The [C/Fe]-[Fe/H] trend alone cannot distinguish metallicity-dependent CCSN yields from delayed AGB C-production.}
    \label{fig:sims_fe}
\end{figure}

Fig.~\ref{fig:sims_fe} compares the low-$\alpha$ [C/Fe]-[Fe/H] trends for different CCSN yield slopes and AGB yield fractions. This trend is predominantly determined by the metallicity dependence of the C yield, similar to the \caah{} trend. Increasing the AGB star contribution ($\fagb$) and the metallicity-dependence of the CCSN yield ($\zcc$) simultaneously leaves the [C/Fe]-[Fe/H] trend unchanged. 
$\fagb$ and $\zcc$ therefore cannot be inferred confidently from the [C/Fe]-[Fe/H] trend alone. This degeneracy between delayed production and metallicity dependent yields is our motivation for including Mg abundances in our sample. This additional information allows us to isolate these two effects more confidently through comparisons to the [Mg/Fe] ratio.

\begin{figure}
\centering
\includegraphics{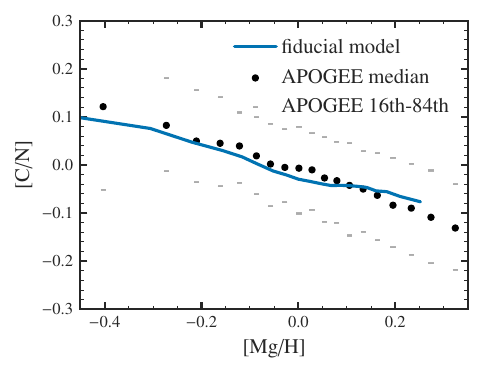}

\caption[]{The median low-$\alpha$ [C/N] ratio as a function of [Mg/H] for our fiducial model (blue line), compared with APOGEE subgiants (medians as black points with 16th and 84th percentile ranges as grey bars). Stars are binned into 20 equal-number bins. Combining our results with the suggested N yield from \citet{james+23} \ (rescaled to our adopted yields) explains the thin-disk evolution of both C and N. 
}
\label{fig:nitrogen}
\end{figure}

N production, also affected by the CNO cycle, is closely tied to C.
This paper builds upon \citet{james+23}, who used similar methods to constrain N yields. As an additional test of both yield models, we combine a rescaled version of their recommended N yields (see discussion in Section~\ref{sec:vice}) with our fiducial choice of C yields. The rescaling is needed because we assume a lower $\Ymg$ and our sample has a higher median [N/Mg] than in \citet{james+23}.
Fig.~\ref{fig:nitrogen} shows [C/N] versus [Mg/H] as predicted by our fiducial model. We are indeed able to closely match the observed [C/N]-[Mg/H] trend.

\subsection{AGB yield modifications}\label{sec:agb_yield_shifts}

\begin{figure}
    \includegraphics{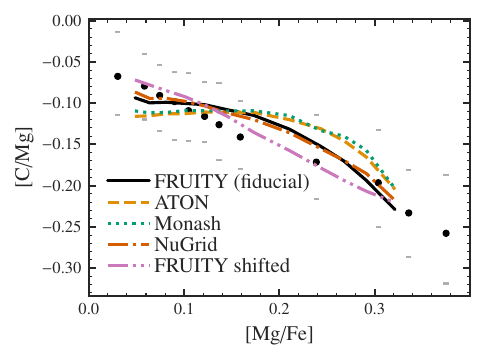}
    
    \caption[]{
        The best-fitting yield combination for each AGB table and the \fruity{} table with a mass shift of 0.7. We fit $\aagb$, $\yo$, and $\zeta$ using the method in Appendix~\ref{sec:extra_mcmc}.
    }
    \label{fig:sims_agb}
\end{figure}

All AGB yield tables predict a similarly shaped trend in the \caafe{} abundance space. Fig.~\ref{fig:sims_agb} shows the best-fitting yield combinations for each AGB table we consider (solving for $\aagb$, $\yo$, and $\zcc$ in Appendix \ref{sec:extra_mcmc}). Although the individual DTDs of each AGB table appear substantially different (Fig.~\ref{fig:agb_yields}), the predicted abundance trends do not show substantial differences. 
This outcome arises because each of the AGB models predicts characteristic delay times for C production of order ${\sim}1$ Gyr, similar to the SN Ia production of Fe (see Fig.~\ref{fig:agb_yields}), leading to a quasi-linear [C/Mg]-[Mg/Fe] relation in all cases. 

However, all AGB tables predict a slightly concave down trend because C production by single stellar populations slows down significantly after a few Gyr, while Fe production through SNe Ia continues up to ${\sim}10$ Gyr (see discussion in Section~\ref{sec:agb}). Our subgiant sample does not exhibit this level of concavity. These discrepancies are significant because Fig.~\ref{fig:sims_agb} plots the mean trend, which is precisely determined through the large sample size. One of the largest discrepancies between our models and data in the left panel is that \aton{} and \monash{} predict an almost flat \caafe{} trend for ${\rm [Mg/Fe]} \lesssim 0.1$.

{These discrepancies in the [C/Mg]-[Mg/Fe] trend indicate that C production by single stellar populations may be more extended than our yield models predict.
We explore this possibility further in} Fig.~\ref{fig:sims_agb}, which also shows a variation of our fiducial model in which we artificially multiply the AGB star progenitor mass a factor of 0.7. Shifting the C production toward lower mass progenitors makes the DTD more extended, comparable to SN Ia. As a result, the mass-shifted \fruity{} model has a more linear \caafe{} trend, better matching the observed abundance pattern. We explore mass-shift variations further in Appendix~\ref{sec:extra_models}.
Adjusting our SN Ia DTD to a prescription that declines more rapidly with stellar population age would also linearize the [C/Mg]-[Mg/Fe] relation.
However, recent GCE models indicate that abundance trends favour an extended DTD  as used here (\citealt{palicio+2024} and \citealt{dubay+24}).

\subsection{The best-fitting yield combinations}

\begin{figure}
    \centering
    \includegraphics{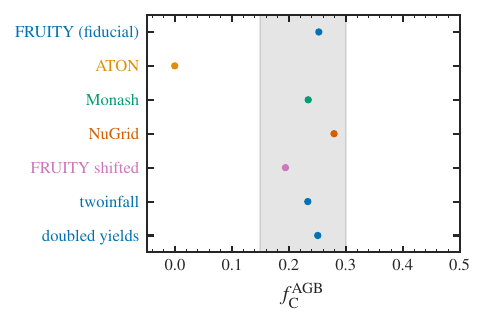}
    
    \caption[]{
     The best-fitting $\fagb$ from the MCMC analysis samples for several model (see Appendix~\ref{sec:extra_mcmc}).We include results for different AGB tables (\fruity{}, \aton{}, \monash{}, \nugrid{}), a model where the \fruity{} is artificially shifted in progenitor mass by a factor of 0.7, a model adopting a different, two-infall star formation history (from Appendix~\ref{sec:extra_models}), and a model where all yields are doubled and outflows adjusted according to Eq.~\ref{eq:mass_loading}. The shaded region represents our suggested $\fagb$ for the Milky Way's thin disk at solar metallicity.
      
    }
    
    \label{fig:mcmc_fagb}
\end{figure}

To derive more quantitative constraints on the origin of carbon, Appendix~\ref{sec:extra_mcmc} presents a yield-fitting framework where we use Monte-Carlo Markov Chains to find the combination of CCSN and AGB yields that match the C abundance trends as binned functions of [Mg/H] and [Mg/Fe]. 
Fig.~\ref{fig:mcmc_fagb} plots the best-fitting $\fagb$ for a variety of models, including alternate AGB yields, star formation histories, and yield scales. \aton{} is the sole outlier, with $\fagb\approx0$, because the AGB yield itself is nearly zero at solar metallicity (see Fig.~\ref{fig:y_agb}).
In all other cases, the best-fitting value of $\fagb$ falls between 0.15 and 0.3 (the shaded region). 

Appendix~\ref{sec:extra_models} explores additional models varying the yield scale, star formation history, and Type Ia supernova scale. Changes to the star formation history or yield scale only slightly change the abundance trends. Instead, the Type Ia SN Fe production directly impacts the derived carbon abundance, with derived $\fagb$ potentially reaching 50\% for high SN Ia Fe yields. However for our best-fitting (mass-shifted) AGB model, a reasonable increase in the Type Ia supernovae rate increases the derived $\fagb$ from 20\% to 25\%. 

\section{Discussion \& Conclusions}\label{sec:conclusions}

Building on the constraints on N production by \citet{james+23}, we quantify the impact of C yield assumptions on Milky Way GCE models. We use \citepos{jack} sample of APOGEE subgiants as our primary observational benchmark, as subgiant atmospheres best reflect their birth C abundances among spectral types (see discussion in Section~\ref{sec:data_selection}).
In our fiducial model, the evolution of [C/Mg] initially increases with time due to the metallicity-dependent C yields from massive stars. Later, AGB stars contribute additional C, sharpening the rise in [C/Mg] with metallicity. 

The \caah{} relation in low-$\alpha$ stars arises as a superposition of the endpoints of evolutionary sequences traced out by different Galactic regions (see Fig.~\ref{fig:c_evo}). As a result, the \caah{} trend is most strongly tied to the total C yield (relative to Mg) at each metallicity [Mg/H]. All AGB star models we consider predict a declining C yield with increasing metallicity. The observed increase in [C/Mg] with [Mg/H], however, implies that the total C yield increases with metallicity. This trend favours a scenario in which massive star yields of C increase with metallicity to compensate for the declining AGB star yields. This outcome could arise due to the heightened metallicity-dependent mass-loss in the rotating models of \citet{LC18}. C and Mg alone, however, do not provide substantial information on the relative contributions of prompt versus delayed production by high- and low-mass stars.
Instead, the \caafe{} relation at fixed [Mg/H] is sensitive to the amount of delayed C production, due to the sensitivity of [Mg/Fe] to delayed SN Ia nucleosynthesis.

From the combined constraints on the [C/Mg] trend with [Mg/H] and [Mg/Fe], we have inferred possible combinations of yields that reproduce thin-disk abundance trends. In all cases, we find that $\Yct / \Ymg \approx 4.2 \pm 0.2$ at solar metallicity, with total C production increasing with increasing metallicity. We also show that low-intermediate mass AGB stars likely produce a fraction of between 0.15 and 0.30 of solar C, across GCE assumptions and AGB yield tables.
At the normalization of stellar yields recommended by \citep{david_fe},  most AGB star models predict C yields within a factor of ${\sim}2$ of our inference.  We find that the best-fitting models require between ${\sim}0.6$ and 2 times more C than predicted by the chosen AGB table.

A range of choices can fit the subgiant data modelled here, but one reasonable model based on our fit results is CCSN yields following Eq.~\ref{eq:y_cc} with $\yo=19.7\times10^{-4}$ and slope $\zcc=8.6\times10^{-4}$ and the \fruity{} AGB yields multiplied by a scaling factor $\aagb=2.14$. If the Mg yield $\Ymg$ is increased relative to our adopted value of $\Ymg=0.97\,Z_{\rm Mg, \odot}$, then all three C yield parameters should be boosted by the same factor.

As a consequence of massive star enrichment dominating the C production in our models, our conclusions are relatively insensitive to the choice of AGB yield model. Of the yield tables tested here, \fruity{} and \nugrid{} provide the best reproduction of the \caafe{} trend. The production of C by AGB stars is too rapid with the \monash{} and \aton{} yields. However, all AGB yield tables predict a \caafe{} relation that is concave down, while the observed relation is linear. Based on this discrepancy, C may be more efficiently produced in lower mass AGB stars than predicted by stellar models, which would increase characteristic delay times and linearize the \caafe{} trend. This discrepancy could also be resolved by a less extended DTD for SN Ia, but such an adjustment would be in tension with recent GCE models \citep{palicio+2024, dubay+24}.

Comparing our models with measurements of C abundances in the gas phase available in the literature reveals a potential shortcoming of our C yield prescription in the ${\rm [O/H]} \lesssim -1$ regime (see Appendix~\ref{sec:extra_gas}). \Hii{} regions at these metallicities typically have $\co \approx -0.6$ but with substantial scatter (see Fig.~\ref{fig:gas_phase}). Our models predict $\co \approx -0.5$ at low metallicity, struggling to explain the lowest [C/O] measurements. Dwarf galaxy GCE models fare even worse, indicating that different galactic environments are not responsible for the difference. Instead, either a larger AGB contribution to C is required, disfavoured by the subgiant trends, or a complex, non-monotonic massive-star production of C is needed. Letting the massive star C yield vary non-linearly with metallicity indeed improves agreement with all of the data that we have considered in this paper. 

As in \citet{james+23}, our models constrain relative as opposed to absolute yields due to the degeneracy between the overall scale and the strength of outflows \citep[e.g.,][]{hartwick1976,james+23, sandford+24}. Rescaling both parameters by similar multiplicative factors leaves the predicted \caah{} relation largely unchanged, to first order (see Appendix~\ref{sec:extra_models}). Breaking this degeneracy requires either age measurements to pin down enrichment time-scales \citep{james+24}, chemical species produced during Big Bang Nucleosynthesis present in accreting gas \citep{cooke+2022, james+25}, or independent constraints from observations of SN \citep[e.g.,][]{rodriguez+23,david_fe}. However, our constraints on the ratio of C/Mg yields, its metallicity dependence, and the AGB star contribution to C should be robust.

As we constrain the yield ratio, variations in the Mg yield may help explain the [C/Mg] trends. $\alpha$-elements like Mg and O, produced primarily in CCSN, are thought to have mostly metallicity-independent yields \citep{andrews+17}. Observationally, the [O/Mg] ratio is approximately solar across most of APOGEE, consistent with this expectation  \citep[][]{weinberg+19}. Nonetheless, our results do not rule out metallicity-dependent O and Mg production on their own.

Our results in this paper demonstrate the utility of empirically calibrated stellar yields. Due to the sensitivity of metal production to poorly understood stellar evolution processes, such as mass loss and convection, these results provide a useful benchmark for stellar models (see discussion in, e.g., \citealt{gil-pons+2022}). With larger samples and more reliable abundance measurements from upcoming spectroscopic surveys, such as SDSS-V's Milky Way Mapper program \citep{kollmeier+25}, our understanding of metal production and the assembly history of our Galaxy will sharpen.
Observing C across a range of environments at higher precision will allow for yet more detailed understanding of C nucleosynthesis, creating stringent tests for stellar evolution theory and refining our understanding of the chemical evolution of the Universe.

\section*{Acknowledgements}

We thank the referee for their detailed review, which has improved the quality of this paper.
We thank Jennifer Johnson and Falk Herwig for helpful comments and discussion. JWJ acknowledges financial support from a Carnegie Theoretical Astrophysics Center postdoctoral fellowship and an Ohio State University Presidential Fellowship. This work was partly supported by NSF grant AST-2307621.

This paper uses APOGEE data \citep{apogee_instrumentation, apogee17, aspcap}. APOGEE is  part of SDSS-IV \citep{sloan_telescope, sdss_iv_overview, sdss17}.

Software that has contributed to this work included  
\VICE~\citep{JW20, james+21},
\textsc{matplotlib} \citep{matplotlib},
\textsc{scipy} \citep{scipy},
\textsc{IPython} \citep{ipy},
\textsc{pandas} \citep{pandas},
\textsc{numpy} \citep{numpy},
\textsc{astropy} \citep{astropy:2013, astropy:2018, astropy:2022},
and 
\textsc{seaborn} \citep{seaborn}
.
Additionally, we thank \citet{OhioSupercomputerCenter1987} for the use of its facilities for the simulations.

\section*{Data Availability}

Data and code used in this paper are happily available upon request.

\bibliographystyle{mnras}
\bibliography{main}


\appendix

\section{NuGrid yield tables and comparison of AGB yields} \label{sec:extra_agb}

In this Appendix, we present updated calculations of the \citet{battino+19, battino+21} yield tables, describe their incorporation into \VICE, and compare AGB yields in more detail. These models use the NuGrid MPPNP post-processing toolkit on models run using the \textsc{mesa} stellar evolution code \citep{mesa}.

\citet{battino+19, battino+21} update the \citet{ritter+18} yield models with new reaction rates and increased mixing. These changes allow for an improved treatment of {\it s}--process isotopes in their models. However, between these models, the only updates are for masses 2 and $3\,\Mo$ at metallicities $Z=0.01, 0.02, 0.001, 0.002$. 

Because the yield tables included in \citet{battino+19, battino+21} are inconsistent with the \citet{ritter+18} definitions of yields, we recalculate these yields from the publicly available data on \url{https://astrohub.uvic.ca}. 

The total ejected mass of $X$ from an AGB star is given by 
\begin{equation}
    M_{X, \rm ej} = \int Z_{X, \rm surf}(t) \dot{m}(t)\ dt,
\end{equation}
where $Z_{X, \rm surf}$ is the surface abundance of $X$, $\dot m$ is the mass loss rate, and the integral is taken over the entire evolution. 
One complication is that {\sc mesa} stellar evolution stops before the end of the AGB phase due to numerical instabilities during the transition to a planetary nebula. 
As a result, the practical calculation is to combine both the ejected surface mass at each time step and the remaining mass not yet ejected:  
\begin{subequations}
    \begin{align}
        M_{X, \rm ej} &= Z_{\rm X, surf, end} (m_{\rm end} - m_{\rm rem}) \\ 
                      &+ \sum_i \frac{1}{2} (Z_X(t_i) - Z_X(t_{i+1}))\ (m(t_i) - m(t_{i+1})),
    \end{align}
\end{subequations}
where $m_{\rm end}$ and $Z_{X, \rm surf, end}$ are the final mass and surface abundance of $X$ at the end of {\sc mesa} evolution, $m_{\rm rem}$ is the mass remaining in the star at the end of {\sc mesa} evolution, and the sum is taken over all time steps in the {\sc mesa} output. We use the final H-free mass as the remnant mass. 

We use the \citet{ritter+18} tables from $M_{X, \rm ej}$, which we can reproduce from the same methods as above. Finally, the $p_{X}$ yields are calculated from $M_{\rm X, ej}$.

\begin{figure*}
    \centering
 	    \includegraphics[scale=1]{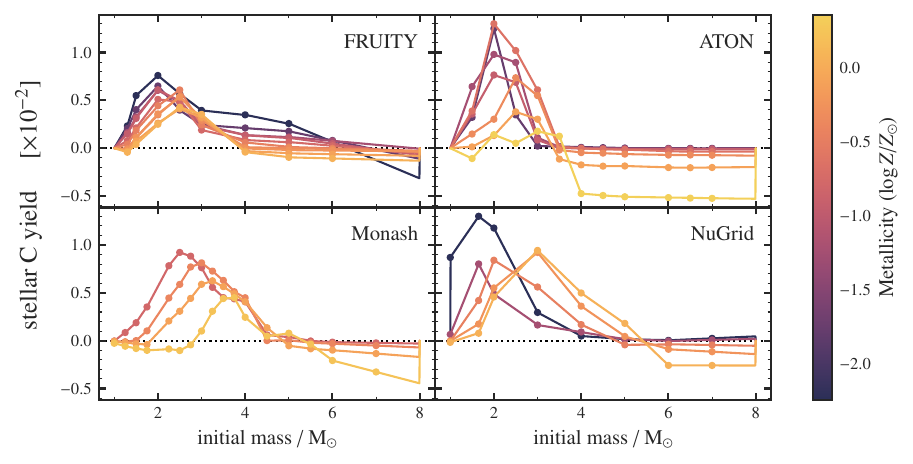}
        \caption[]{The net fractional C yield from AGB stars as a function of initial stellar mass and metallicity (similar to the left panel of Fig.~\ref{fig:agb_yields}). Each panel represents yields from one of four AGB studies from the literature: \fruity{}, \aton{}, \monash{}, \nugrid{} (see Section \ref{sec:agb}).  The black dotted line shows $\Ycagb=0$ for reference. 
        }

        \label{fig:y_agb}
\end{figure*}

Fig.~\ref{fig:y_agb} compares each AGB yield table as a function of both mass and metallicity. 
In general, all tables have a similar qualitative shape. AGB C production peaks at initial masses of $\sim$ 2--4 $\Mo$, and the peak mass increases slightly with metallicity. At higher metallicities, intermediate mass stars ($M \gtrsim 5\Mo$) start to deplete C and most stars produce less C in general. The strongest outlier is the \aton{} model, which predicts that most AGB stars deplete C at solar metallicity. Otherwise, it is reasonable to expect these models of C AGB production to predict similar GCE abundance trends.

\section{Additional results}\label{sec:extra_models}
In this Appendix, we explore additional models not included in the main text. We test modifications to the AGB yield table, the star formation history, and the yield scale. In most cases, the results do not change too substantially.

\subsection{Model parameters and measurement scatter} \label{sec:extra_scatter}

\begin{table}
	\centering
        \caption[]{Description of the yields used for models presented in this
        paper (see Sections \ref{sec:agb}, \ref{sec:vice}, and \ref{sec:results}
        for details).
        }
	\label{tab:model_parameters}

	\begin{tabular}{l l l l l l l l l}
		\hline
            name & $\aagb$ & $\ycagb$ & $\yo/10^{-4}$ & $\zcc/10^{-4}$ \\ 
            \hline
            fiducial & 2.14 & FRUITY & 19.7 & 8.6 \\
            $\zcc=0.0004$ & 2.14 & FRUITY & 19.7 & 4.3 \\
            $\zcc=0.0012$ & 2.14 & FRUITY & 19.7 & 12.8 \\
            $\fagb = 0$ & 0.0 & FRUITY & 27.8 & 8.6 \\
            $\fagb = 0.5$ & 4.11 & FRUITY & 12.8 & 8.6 \\
            lower $\fagb$, $\zcc$ & 0 & FRUITY & 27.8 & 4.3 \\
            higher $\fagb$, $\zcc$ & 4.11 & FRUITY & 12.8 & 12.8 \\
            \aton{} & 1.53 & ATON & 25.1 &  14.8 \\
            \monash{} & 2.04 & Monash & 19.9 & 15.9 \\
            \nugrid{} & 0.82 & NuGrid & 19.7 & 4.7 \\
            \fruity{} shifted & 1.48 & FRUITY($M/0.7$) & 22.1 & 5.3 \\
            mass shift 0.5 & 2.62 & FRUITY($M/0.5$)  & 22.2 & 4.3 \\
            mass shift 0.7 & 1.92 & FRUITY($M/0.7$) & 20.7 & 6.3 \\
            mass shift 1.5 & 2.43 & FRUITY($M/1.5$) & 22.6 & 4.3 \\
		\hline
	\end{tabular}
\end{table}

Table \ref{tab:model_parameters} lists the parameters of the models used in this paper.

To add artificial scatter into our models for use in the model fitting described in Appendix~\ref{sec:extra_mcmc}, we fit polynomials to the reported internal APOGEE error with metallicity: 
\begin{subequations}
\label{eq:uncertainties}
\begin{align}
    \delta {\rm [Mg/H]} &= 0.0652\,x^2 + 0.00522\,x + 0.0338 \\
    \delta {\rm [Mg/Fe]} &= 0.00793\,x^2 - 0.00802\,x + 0.0138 \\
    \delta {\rm [C/Mg]} &= -0.0378\,x + 0.03506,
\end{align}
\end{subequations} 
where $x = {\rm [Fe/H]}$ for brevity. In detail, these polynomials should also vary with surface gravity and effective temperature. However, as all of our stars are subgiants, such effects should be smaller.

\subsection{AGB yield mass shifts}

In Section~\ref{sec:agb_yield_shifts}, 
we noted that shifting the \fruity{} table artificially in progenitor mass improves agreement with observed abundance trends. Here, we show the effect on abundance trends of shifting the \fruity{} table initial masses by different factors. 

{Fig.~\ref{fig:sims_degens} shows variations of our fiducial model in which we artificially multiply the AGB star progenitor mass by factors of $0.5$, $0.7$, and $1.5$.
When the AGB C production is weighted towards more massive AGB stars, the DTD becomes more rapid, leading to higher [C/Mg] at high [Mg/Fe]. On the other hand, shifting the C production toward lower mass progenitors makes the DTD more extended. In Fig.~\ref{fig:sims_degens}, the models with AGB C shifted towards lower progenitor masses reduce or reverse the concavity of the \caafe\ trend. For the extreme case, where the progenitor masses are shifted by a factor of 0.5, the AGB production is more delayed than SN Ia, resulting in a concave up trend. Prefactors around $\sim$$0.7$, corresponding to a slight shift in C production toward lower mass AGB stars, lead to the most linear [C/Mg]-[Mg/Fe] relation as indicated by the data.
While our mass shift models are deliberately simple, they illustrate that the observed APOGEE trends would be better fit if AGB production were shifted towards longer-lived stars.
Adjusting our SN Ia DTD to a prescription that declines more rapidly with stellar population age would also linearize the [C/Mg]-[Mg/Fe] relation.
However, recent GCE models indicate that abundance trends favour the more extended DTD we adopt here (see discussion in \citealt{palicio+2024} and \citealt{dubay+24}).

\subsection{Star formation history and yield scale}

\begin{figure*}
    \includegraphics{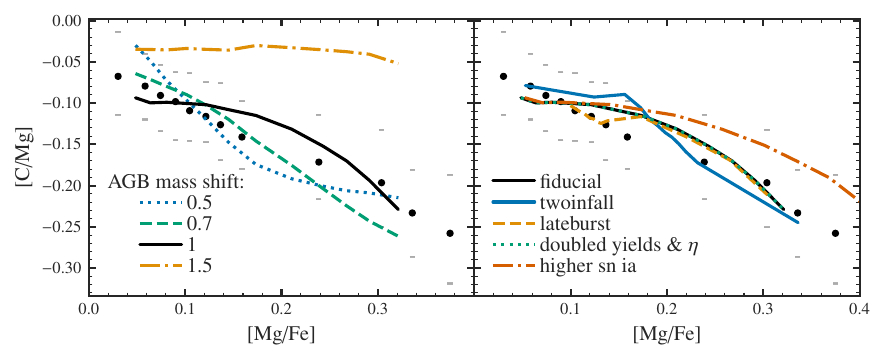}
    
    \caption[]{
        Similar to Fig.~\ref{fig:zeta_f}, but focusing on [C/Mg] versus [$\alpha$/Fe]. Models where the \fruity{} AGB yields are artificially shifted by different multiplicative factors in progenitor mass, but maintaining the same $\fagb$. 
        {\bf Right:} Models with alternative star formation histories
        (Eqs.~\ref{eq:lateburst}, \ref{eq:twoinfall}), and a model where all yields are doubled (with mass-loading adjusted accordingly).
        }
    \label{fig:sims_degens}
\end{figure*}

In the right-hand panel of Fig.~\ref{fig:sims_degens}, we consider two alternative SFHs.
Motivated by the findings of Mor et al. (2019) and Isern (2019; see discussion in \JJ), the \textit{lateburst} model adds a Gaussian-shaped burst to the fiducial inside-out model, 
\begin{equation}\label{eq:lateburst}
    \dot{\Sigma}_\text{lateburst} \propto \dot{\Sigma}_\text{inside-out} \left(1 + A\,e^{-(t-t_{\rm burst})^2/2\sigma^2_{\rm burst}} \right),
\end{equation}
where $A=1.5$ represents the amplitude of the burst, $t_\text{burst}=10.8$\,Gyr is the time where the burst is strongest, and $\sigma_\text{burst}=1$\,Gyr is the width of the burst.

In the right-hand panel of Fig.~\ref{fig:sims_degens}, we consider a two-infall evolutionary history \citep[e.g.,][]{chiappini+97,spitoni19, spitoni21, spitoni+23}. This prescription posits that the majority of the stellar mass of the Galactic disk was built up during two events of substantial metal-poor accretion separated in time. The two-infall scenario offers a natural explanation \citep{spitoni19, spitoni21} for the bimodal distribution of [$\alpha$/Fe] ratios observed in the MW \citep[e.g.,][]{hayden+15}. However, the two-infall model has recently faced challenges when confronted with large catalogues of stellar age measurements \citep{dubay+25}. 
For this, we switch \VICE{} to `infall mode' and let it compute the SFR from a specified accretion history, in contrast to the other models where we specify the SFH (see discussion in Section~\ref{sec:vice}). Our accretion history, following \citet{dubay+24}, is given by
\begin{equation}\label{eq:twoinfall}
\dot{\Sigma}_{\rm in} \propto e^{-t/\tau_1} + A \theta(t - t_{2}) e^{-t/\tau_2},
\end{equation}
where $\tau_1=1\,$Gyr, $\tau_2=4\,$Gyr control the duration of the first and second burst, $t_2=4$\,Gyr specifies the onset of the second infall epoch, $A$ is the relative amplitude of the second burst, and $\theta$ is the Heaviside step function. We chose $A(R)$ such that the ratio between the stellar mass produced in the first and second burst by $13.2\,{\rm Gyr}$ matches the ratio between the thin and thick disks from \citet{BHG16}.

The right panel of Fig.~\ref{fig:sims_degens} compares the fiducial SFH with these lateburst and two-infall variations. The differences in the \caafe\ trend are small. This model predicts a different distribution in [Mg/Fe] as expected \citep[e.g.,][]{JW20}, which we do not show here.
Most importantly, the relationship between these abundance ratios is largely unaffected by the assumed SFH.

Lastly, we demonstrate that our conclusions about yield ratios are unaffected by the assumed scale of stellar yields and mass loading in outflows.
The fiducial yield scale adopted here, from \citet{david_fe}, is relatively low. We require only moderate outflows to reproduce solar metallicity. However, models with higher outflows and higher yields give similar predictions because of the 
strong degeneracy between the normalizations of elemental yields and mass-loading (see discussion in, e.g., \citealt{sandford+24} and Appendix B of \citealt{james+23}). 
If we were to swap mass-loading for inward radial gas flows contained within the Galactic disk, a similar relationship would arise between the yield scale and the speed of the flow \citep{james+25}.
Our parameterization of $\eta$ illustrates this point (see Eq.~\ref{eq:mass_loading}) -- choosing a higher value of $y_{\rm Mg}^{\rm CC}$ will similarly result in higher values of $\eta$ while maintaining the same metallicity profile, $Z_\text{Mg}(R)$. 
In the right panel of Fig.~\ref{fig:sims_degens}, we consider a model in which all yields are doubled (and mass loading approximately doubled by Eq.~\ref{eq:mass_loading}). As expected, the trend in [C/Mg]-[Mg/Fe] is not significantly affected. Based on this degeneracy, our investigations ultimately constrain the \textit{relative} yields of C and Mg, in the same way that Johnson et al. (2023) constrained the relative yields of N and O, as opposed to the total yields of any element.

\section{Yield Inference} \label{sec:extra_mcmc}

In this Appendix, our goal is to convert the qualitative understanding described up to this point into more quantitative constraints on C yields. 
In Section~\ref{sec:results}, we showed that the \caah{} trend reflects the overall metallicity dependence of the total C yield from single stellar populations, while the \caafe{} trend reflects the amount of delayed C and its DTD. Here, we construct Monte-Carlo Markov chain (MCMC) models to explore possible abundance combinations matching the observed trends. From these models, we can derive the most likely C yields given current observations.

\subsection{Monte-Carlo Markov Chain yield inference} \label{sec:mcmc_methods}

To efficiently explore a wide range of yield combinations, we use an approach that we refer to as {\it process tracking}. In GCE, total elemental abundances are (to first order) linear with respect to combinations of yields \citep[e.g.,][]{WAF17}. 
Process tracking exploits this linearity to integrate the CCSN and AGB yields separately, later expressing the C abundance as a linear combination of the two.
We set up \VICE {} to integrate each process for C as different chemical elements.
The C abundances then follow as $Z_\text{C,tot} = \sum a_i Z_i$, where $a_i$ and $Z_i$ are the amplitude and abundance evolution of the $i$th process. 

We emphasize that there are some limitations to the process-tracking approximation. Capturing the effects of many other GCE parameters -- such as the star formation history, outflow, and radial migration prescriptions --  requires separate multi-zone integrations for each parameter variation. 
Metallicity-dependent yields also introduce a second-order effect (metallicity depends on total yields, which influence the metallicity), but this effect should be insignificant for small changes to total yields.

Our first step is to run a multi-zone model tracking the abundances of each process we are interested in (in this case, CCSN and AGB star products). 
To use all the information available for [C/Mg] trends with [Mg/Fe] and [Mg/H], we calculate the mean $Z_\text{C} / Z_\text{Mg}$ for each 2D bin in [Mg/Fe] and [Mg/H] for both the data and the model. We add Gaussian-distributed measurement uncertainties to the stars in the model as described in the Appendix~\ref{sec:extra_scatter}.  We use bins with a constant width of 0.04 dex, ranging from ${\rm [Mg/H]}=[-0.5, 0.5]$ and ${\rm [Mg/Fe]} = [-0.1, 0.4]$. We remove 2D bins with fewer than 3 stars in either the model or data.
We describe C yields with three different parameters:
\begin{enumerate}
    \item $\aagb$: Prefactor for an AGB yield table
    \item $\yo$: Constant CCSN yield
    \item $\zcc$: Linear CCSN yield ($y = \zcc (Z/\Zo)$),
\end{enumerate}
where $\yo$ represents the C yield at low metallicity.

We construct a likelihood function based on the difference in the mean $Z_\text{C}/Z_\text{Mg}$ between the data and model in each bin. 
We use the linear as opposed to logarithmic C abundance (i.e., [C/H]) because the linear combination of enrichment channels from process tracking predicts $Z_\text{C}$. 
Given the predicted abundance for the model in a given bin $b$,  $Z_{\rm C, \it b}^{\rm model}$, the associated observed abundance, $Z_{\rm C, \it b}^{\rm obs}$, and the error on the mean $\sigma_b$, the (Gaussian) log likelihood is given by
\begin{equation} \label{eq:mcmc_likelihood}
    \log {\cal L} = \sum_b -\ln(\sigma_{b}) -\left(\frac{Z_{\rm C, \it b}^{\rm model} - Z_{{\rm C}, b}^{\rm obs}}{2\sigma_{b}}\right)^2 - \frac{1}{2}\log(2\pi).
\end{equation}
We calculate the total variance in each bin from the standard error on the mean of the observations in a bin $\sigma_{b, \rm obs}$, the standard error on the mean for each model component $i$, and $\sigma_{b, i,\rm model}$. Altogether, 
\begin{equation}
    \sigma_{b}^2 = \sigma_{b,\rm\ obs}^2  +  \left(\sum_{i} a_i\, \sigma_{b,i,\ \rm model}\right)^2
\end{equation}
where $a_i$ is the associated coefficient ($\aagb$, $\yo$, $\zcc$) for each yield process.
We add the model standard deviations linearly since the C abundances are calculated from the same sample for each process.

We adopt the following priors for each model parameter: $\aagb \sim N(1, 1)$, $\zcc \sim N(0, 10^{-3})$, and $\yo-\zcc \sim N(2\times10^{-3}, 1\times10^{-3})$, where $N(\mu, \sigma)$ is a Gaussian distribution with mean $\mu$ and standard deviation $\sigma$. (For the doubled yields and outflows model, the priors are correspondingly doubled). 
We run the MCMC using Turing.jl \citep{turingjl} with a No-U-turn sampler (an adaptive Hamilton Monte-Carlo algorithm, \citealt{NUTS}) with an acceptance rate of 0.65. We combine 16 independent chains with 3,000 steps each for a total of 48,000 samples.
The median parameters derived from this model provide the yield choices for the AGB yield models shown in the left panel of Fig.~\ref{fig:sims_degens}.

\subsection{Yield inference results} \label{sec:mcmc_results}

Based on the procedure above, we derive our best-fitting yield set (not the fiducial yields described in Section~\ref{sec:agb}). For the \fruity{} AGB yield set, we find $\aagb = 1.55$, $\yo =22.3\times10^{-4}$, and $\zcc=7.9\times10^{-4}$. Relative to our fiducial model, $\aagb$ is lower, but the normalization $\yo$ of the CCSN yield is higher. For the model where \fruity{} is shifted by a mass factor of 0.7, we instead find $\aagb = 2.34$, $\yo = 19.7\times 10^{-4}$ and $\zcc=3.2\times 10^{-4}$.  The fiducial yields enforce a low-metallicity [C/O] plateau of \about{-0.5} for better agreement in Section~\ref{sec:extra_gas}. We present the best-fitting parameters for each model or yield set considered here in Appendix~\ref{sec:extra_mcmc}. 

The left panel of Fig.~\ref{fig:mcmc_ytot} shows the inferred total C yield for each AGB yield table, along with the \fruity{} table shifted in mass by a prefactor of 0.7. While there is some variability, all models predict  a dynamic range of $y_\text{C}^\text{tot} / y_\text{Mg}$ of approximately $\sim$$0.5$. Therefore, the choice of AGB yield table only slightly affects the inferred total yield.

The middle panel of Fig.~\ref{fig:mcmc_ytot} shows the inferred AGB fraction (Eq.~\ref{eq:f_agb}) at solar metallicity for each model (see also Table~\ref{tab:mcmc_results}). 
In short, all of our yield models favour a scenario in which AGB stars are responsible for less than half, and likely near 20 per cent, of the C present in near-solar metallicity environments. No model predicts a median $\fagb > 0.3$ and the 98th percentile upper-limits are all below $\fagb=0.4$. All models except \aton{} predict that at least $\approx 8$ per cent of the solar C comes from AGB stars (from the 2nd percentile lower-limits). The SFH and overall yield scale have little impact on the inferred $\fagb$.  

The right panel of Fig.~\ref{fig:mcmc_ytot} compares the inferred scaling of AGB yields, $\aagb$. For most tables, we find $\aagb$ near unity, indicating accurate total C yields from AGB stars.
Doubling our yields requires higher values of $\aagb$ to produce more C at fixed $\fagb$.
For \monash{} and \aton{}, $\aagb$ is almost exactly $1$. For \aton{}, $\aagb=1$ is not unreasonable despite producing little C at solar metallicity because \aton{} produces similar C as other models at lower metallicities. 
However, these models struggle to reproduce the [C/Mg]-[Mg/Fe] relation, which suggests issues related to the mass-dependence of the yields (see discussion in Section \ref{sec:agb_results}).

Finally, the choice of the Fe SN Ia DTD directly impacts our conclusions.  From the right panel of Fig.~\ref{fig:mcmc_ytot}, we note that increasing the SN Ia Fe yield by a factor of 1.2 (and decreasing the CC Fe yield to maintain the same total Fe production) results in a similar increase (${\sim}1.2\times$) in the inferred $\fagb$ . Increasing the SN Ia contribution to Fe stretches the trends in [Mg/Fe], causing [C/Mg] to flatten with [Mg/Fe] unless the delayed C yield is also increased. In this framework, constraints on delayed C contributions are limited by our understanding of delayed Fe production.

Fig.~\ref{fig:mcmc_ytot} shows the posterior distributions of our MCMC fit to the C yield models assuming the \fruity{} AGB table. Each parameter is relatively well-constrained. The most prominent degeneracy, between $\yo$ and $\aagb$, represents the constraint on the equilibrium abundance of C near solar. 

Table~\ref{tab:mcmc_results} provides the best-fit median values and quantile-based uncertainties for each parameter for each AGB yield table and GCE model discussed here. 

\begin{table*}
    \caption{
        Best-fit parameters (medians and 16-84th percentile ranges) from our MCMC fit for each model discussed.
        The columns are:  the model name, best-fitting log posterior probability $\log p$, AGB scaling $\aagb$, CCSN low-metallicity yield $\yl$, the CCSN yield slope $\zcc$, the AGB fraction $\fagb$, and the total C yield at solar metallicity $\Yct$. 
    }
    \label{tab:mcmc_results}

{ 
\renewcommand{\baselinestretch}{1.2}
    \begin{tabular}{l l l l l l l l l l l} 
    \hline
    model            
    & $\log p$ & $\aagb$ & $\yo/10^{-4}$ & $\zcc/10^{-4}$ &  $\fagb$ & $\Yct/10^{-4}$\\    %
\hline
\fruity          &  -536.38 & $2.15\pm0.04$  &  $19.75\pm0.15$  &  $8.6\pm0.1$  &  $0.253\pm0.005$  &  $26.42\pm0.04$\\ 
\aton            &  -657.17 & $1.53\pm0.03$  &  $25.1\pm0.06$  &  $14.79\pm0.2$  &  $-(5.5\pm0.1)\times10^{-4}$  &  $25.09\pm0.06$\\ 
\monash          &  -871.48 & $2.05\pm0.05$  &  $19.88\pm0.2$  &  $15.9\pm0.3$  &  $0.234\pm0.006$  &  $25.95\pm0.06$\\ 
\nugrid          &  -847.15 & $0.827\pm0.016$  &  $19.73\pm0.16$  &  $4.7\pm0.1$  &  $0.279\pm0.006$  &  $27.38\pm0.03$\\ 
\fruity\ shifted    &  -307.60 & $1.48\pm0.02$  &  $22.11\pm0.09$  &  $5.29\pm0.09$  &  $0.194\pm0.003$  &  $27.44\pm0.03$\\ 
doubled yields (and $\eta$) &  -551.30 & $4.26\pm0.08$  &  $39.6\pm0.3$  &  $16.95\pm0.19$  &  $0.251\pm0.005$  &  $52.84\pm0.08$\\ 
lateburst        &  -386.87 & $2.45\pm0.05$  &  $18.55\pm0.2$  &  $7.41\pm0.14$  &  $0.291\pm0.007$  &  $26.16\pm0.05$\\ 
two-infall        &  -436.15 & $1.99\pm0.03$  &  $20.29\pm0.12$  &  $8.26\pm0.1$  &  $0.233\pm0.004$  &  $26.47\pm0.04$\\ 
higher SN Ia     &  -392.76 & $3.86\pm0.06$  &  $13.2\pm0.2$  &  $12.09\pm0.13$  &  $0.477\pm0.009$  &  $25.17\pm0.05$\\ 
higher SN Ia \& \fruity\ shifted &  -283.12 & $1.79\pm0.03$  &  $20.6\pm0.1$  &  $5.96\pm0.09$  &  $0.238\pm0.003$  &  $27.02\pm0.03$\\
\hline
\end{tabular}

}
    
\end{table*}

\begin{figure}
\centering
\includegraphics{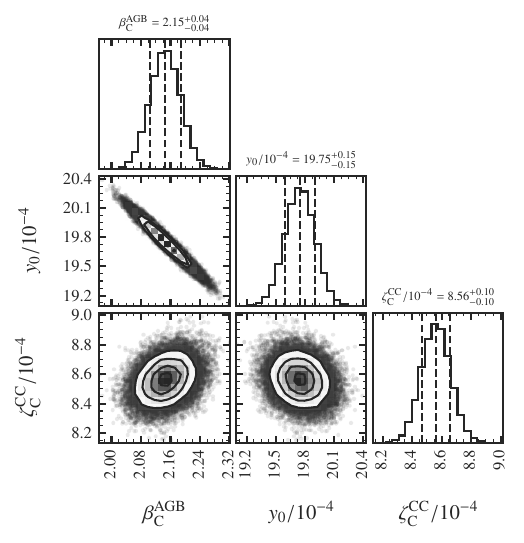}
\caption[]{
The posterior distributions from the MCMC simulation for the analytic model fit to the mean abundance trends (see Appendix~\ref{sec:mcmc_methods}).
$\aagb$ represents the scale factor for the AGB yields, and $\yo$, $\zcc$ represent the constant and linear components of the CCSN yield. 
}

\label{fig:mcmc}
\end{figure}

\begin{figure}
\centering
\includegraphics{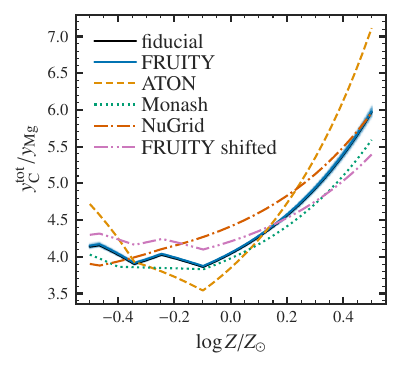}
\caption[]{
The population-averaged yield ratio, $\Yct / \Ymg$, based on our MCMC fits to the [C/Mg]-[Mg/H]-[Mg/Fe] relation. We also show individual
samples from the FRUITY MCMC model
}
\label{fig:mcmc_ytot}
\end{figure}

\section{Gas-Phase Abundances}\label{sec:extra_gas}

\begin{figure*} 
\centering
\includegraphics[]{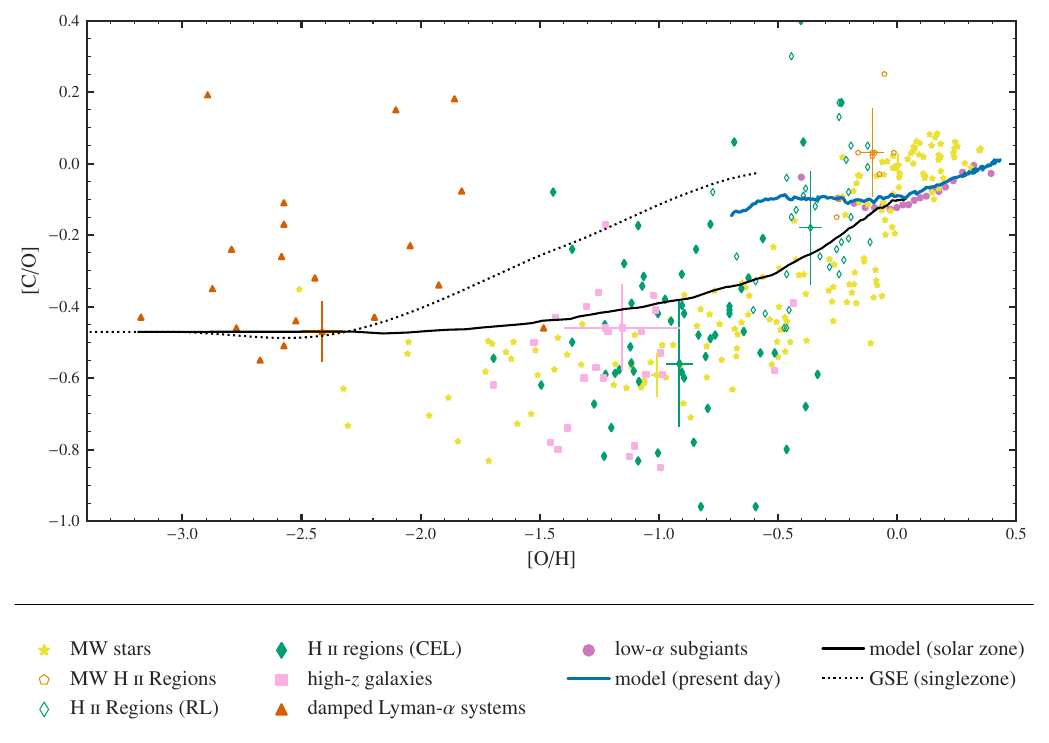}
\caption[]{
Gas-phase C abundances. 
We plot the present-day gas-phase abundances predicted by our fiducial model as a thick blue line and the evolution of the $R=8$ kpc zone as a solid black line.
The black dotted line is a simple, one-zone model for a GSE-like dwarf galaxy.
Points represent measurements for MW stars (yellow stars), MW \Hii{} regions (recombination lines, orange pentagons), extragalactic \Hii{} regions from recombination lines (green empty diamonds), extragalactic \Hii{} regions from collisional excited lines (green full diamonds), high redshift galaxies (pink squares), damped Lyman-$\alpha$ systems (orange triangles),  and the low-$\alpha$ APOGEE subgiant sample we used here (purple circles). 
The points with error bars show representative median uncertainties for their associated population. Six measurements based on collisional excited lines or in damped Lyman-$\alpha$ systems (with large uncertainties) fall above or below the axis limits.

Data for Milky Way stars are from \citet[][]{amarsi+19},
MW \Hii{} regions from \citet[][]{mendez-delgado+22, esteban+05, esteban+13}, 
extragalactic recombination line \Hii{} regions from \citet[][]{peimbert+05, skillman+20, toribio-san-cipriano+16, toribio-san-cipriano+17, esteban+14, esteban+09}, 
extragalactic collisional excited lines \Hii{} regions from \citet[][]{garnett+95, senchyna+17, izotov+thuan99, garnett+99, berg+16, berg+19, pena-guerrero+17}, 
high redshift galaxies from \citet[][]{steidel+16, stark+14, matthee+21, mainali+20, jones+23, amorin+17, iani+23, james+14, erb+10, bayliss+14, berg+18, christensen+12, arellano-cordova+2022}, and
damped Lyman-$\alpha$ systems from \citet[][]{omera+01, cooke+14, welsh+22, ellison+10, welsh+20, cooke+18, riemer-sorensen+17, DZ+03, cooke+17, cooke+11, dutta+14, morrison+16, srianand+10, pettini+08, kislitsyn+24, cooke+15}.
}

\label{fig:gas_phase}
\end{figure*}

As an additional test of our model, we compare the predictions against C abundances
in the gas phase. We consider measurements in Galactic and extragalactic \Hii{} regions, Damped Lyman-$\alpha$ systems,\footnote{Damped Lyman-$\alpha$ systems (or DLAs) are gas clouds observed as absorption spectra in front of background quasars.} and Milky Way halo main-sequence stars (all references in the caption of Fig.~\ref{fig:gas_phase}, which we do not repeat here for brevity).
While we used Mg as our representative alpha element in our APOGEE sample, we shift focus to O in this Appendix because O is more readily observed in the gas phase.
C abundances are challenging to measure in \Hii{} regions because of a lack of strong recombination lines. The collisionally excited lines of C fall in the far ultraviolet without nearby reference H lines \citep[][]{skillman+20}. 
Metals may also be trapped in dust grains or in unobserved ionization states, requiring model-dependent corrections \citep[e.g.,][]{MM19}.
We will see below that, despite these challenges, a global trend in [C/O] with metallicity nevertheless emerges.

Fig.~\ref{fig:gas_phase} compiles literature measurements of [C/O]. The \Hii{} regions span a wide range of host galaxies: from the Milky Way and nearby spiral galaxies, to blue compact dwarf galaxies, to high-redshift star-forming and dwarf galaxies. We also include a sample of thick disk and halo stars for reference. 
Near solar metallicity, most measurements have ${\rm [C/O]} \approx 0$.
Below solar metallicity, [C/O] appears to follow a non-monotonic trend, reaching a minimum of $\co  \approx -0.6$ around ${\rm [O/H]} \approx -1$, and possibly increasing again at the very lowest metallicities.

To better represent the evolution of low-metallicity and dwarf galaxy environments, we construct an example {\it Gaia}-Sausage Enceladus (GSE, \citealt{meza+05, belokurov+18, helmi18}) --like model. This model has a single-zone,  evolving one homogeneous gas supply without radial migration. Compared to our Milky Way model, the GSE-like model uses the same yields but has a lower star-formation efficiency, a more rapid decline in star formation over time, and significantly higher outflows.
Our GSE-like model parameters are derived from the best-fitting parameters of \citet{james_dwarfs}, except adjusting outflows to our yield scale. We adopt a constant mass loading of $\eta=5.74$, total star formation time of 10.73 Gyr, a constant star formation efficiency time-scales of 26.60 Gyr, and an exponential infall history $\dot \Sigma_{\rm in} = \exp(-t / 2.18\,{\rm Gyr})$ (in contrast to star formation rate specification for our fiducial model). We adopt the same \citet{kroupa01} initial mass function as the fiducial model. Fig.~5 from \citet{james_dwarfs} shows the evolution of [Mg/Fe]-[Fe/H] of this model.

The gas-phase C abundances reveal a breakdown of our fiducial yield prescription (presented in Section~\ref{sec:nucleosynthesis}). While the fiducial model sits well within the scatter of C/O ratios observed in low-redshift \Hii{} regions near solar metallicity, the evolution of the $R = 8$ kpc annulus overestimates the mean [C/O] at lower metallicities. The CCSN-only, low-metallicity plateau of [C/O] in the fiducial model is near $-0.5$, whereas the data sit closer to [C/O] $\sim -0.7$ near [O/H] $\sim - 1$. Furthermore, our GSE-like model overpredicts [C/O] across most of this range. Due to the higher mass-loading factor, this evolutionary track can be interpreted as shifted toward low metallicity relative to the solar zone at $R = 8$ kpc. Consequently, near-solar [C/O] ratios arise at low [O/H], resulting in a relatively poor fit to the gas-phase data. We note that our subgiant sample only extends as low as a metallicity of ${\rm [O/H]} \sim -0.4$, so any application of the model at these metallicities requires substantial extrapolation. 

One possible remedy would be to decrease our assumed minimum CCSN yield $y_{\rm low} = \yo - \zcc$, which would lower the [C/O] plateau and improve agreement with observations. A lower $y_{\rm low}$ also requires a higher $\fagb{}$ to reach the observed [C/Mg] near solar metallicity. However, a high $\fagb{}$ model would be in tension with the subgiant abundance trends (see Fig.~\ref{fig:agb_yields}). 
Based on these opposing indications of the value of $\fagb$ from our subgiant sample and the gas-phase data in Fig.~\ref{fig:gas_phase}, it is difficult to reconcile the two observations in our models.

Another possibility is that the massive star production of C (or O) exhibits a complex metallicity dependence. The increasing [C/O] abundances at the very lowest metallicities as observed in damped Lyman-$\alpha$ systems already hint at a more nuanced C nucleosynthetic landscape. Given the abundance trends observed in Fig.~\ref{fig:gas_phase}, we propose the following possibility for C production. (1) At the very lowest metallicities, massive stars may produce copious amounts of C through rotation, alternative IMF, or differing supernova mechanisms. 
This would now be observed as CEMP-no stars \citep[e.g., review and references in][]{FN15} and the elevated C abundances of damped Lyman-$\alpha$ systems. (2) C production sharply drops as metallicity increases, reaching a minimum between [O/H] of -3 and -1.5. Stars and (dwarf) galaxies with the lowest C/O ratios would be formed in this stage. (3) As metallicity increases, massive star C yields steeply increase. In addition, C production from delayed AGB stars becomes more and more important, accelerating an increase of C abundances relative to O on the approach to solar metallicity. (4) Finally, in the regime of our own galaxy, the rapid increase of CCSN production of C may temper, procuring the gently sloping trend apparent in subgiants today.

\citet{berg+19} presented a suite of GCE models that complement our investigations in this section.
They focused on the effects of the bursty SFHs that dwarf galaxies are thought to have experienced \citep[e.g.][]{mcquinn+10} for a given yield prescription.
Our approach is the inverse; we have determined the effects of different choices of stellar yields with fewer considerations of the detailed SFH.
Our models indicate that the exact choice of SFH is inconsequential to the subgiant abundance trends in APOGEE, primarily because our models predict a chemical equilibrium to arise early in the thin disk epoch.
Advocated for by \citet{james+24} based on age-metallicity trends, the equilibrium scenario suggests that the abundance trends in our subgiant sample should reflect trends in C, Mg, and Fe yields (see Section~\ref{sec:results}).
However, sudden bursts of star formation should always lead to chemical perturbations on some level, even in a state of equilibrium \citep{JW20}.
Observing different galaxies at different phases in their burst cycles should drive substantial abundance scatter, as the models in \citet{berg+19} indicate.
Adjustments to the C yields, as we have investigated, would instead shift all galaxies uniformly in the [C/O]-[O/H] plane.

\bsp	
\label{lastpage}
\end{document}